\documentclass[aps,pra,twocolumn,amsmath,amssymb,tightenlines,epsfig,floatfix,superscriptaddress]{revtex4-1}
\usepackage{graphicx}
\usepackage{dcolumn}	
\usepackage{bm}			
\usepackage{amsfonts}
\usepackage{xspace}
\usepackage{color}
\usepackage{epstopdf}
\usepackage{multirow}
\usepackage{amsmath}
\usepackage{float}
\usepackage{hyperref}
\usepackage{cleveref}
\usepackage{amsmath}
\usepackage{mathtools}
\usepackage[section]{placeins}
\begin{document}
\title{Multi-parameter estimation with multi-mode Ramsey interferometry}
\author{Xinwei Li}
\author{Jia-Hao Cao}
\author{Qi Liu}
\affiliation{State Key Laboratory of Low Dimensional Quantum Physics, Department of Physics, Tsinghua University, Beijing 100084, China}
\author{Meng Khoon Tey}
\email{mengkhoon\_tey@mail.tsinghua.edu.cn}
\affiliation{State Key Laboratory of Low Dimensional Quantum Physics, Department of Physics, Tsinghua University, Beijing 100084, China}
\affiliation{Frontier Science Center for  Quantum Information, Beijing 100084, China}
\affiliation{Collaborative Innovation Center of Quantum Matter, Beijing 100084, China}
\author{Li You}
\email{lyou@mail.tsinghua.edu.cn}
\affiliation{State Key Laboratory of Low Dimensional Quantum Physics, Department of Physics, Tsinghua University, Beijing 100084, China}
\affiliation{Frontier Science Center for  Quantum Information, Beijing 100084, China}
\affiliation{Collaborative Innovation Center of Quantum Matter, Beijing 100084, China}
\affiliation{Beijing Academy of Quantum Information Sciences, Beijing 100193, China}

\def\clr{\color{red}}


\date{\today}

\begin{abstract}
Estimating multiple parameters simultaneously is of great importance to measurement science and application. For a single parameter, atomic Ramsey interferometry (or equivalently optical Mach-Zehnder interferometry) is capable of providing the precision at the standard quantum limit (SQL) using unentangled probe states as input. In such an interferometer, the first beam splitter represented by unitary transformation $U$ generates a quantum phase sensing superposition state, while the second beam splitter $U^{-1}$ recombines the phase encoded paths to realize interferometric sensing in terms of population measurements. We prove that such an interferometric scheme can be directly generalized to estimation of multiple parameters (associated with commuting generators) to the SQL precision using multi-mode unentangled states, if (but not iff) $U$ is orthogonal, i.e. a unitary transformation with only real matrix elements. We show that such a $U$ can always be constructed experimentally in a simple and scalable manner. The effects of particle number fluctuation and detection noise on such multi-mode interferometry are considered. Our findings offer a simple solution for estimating multiple parameters corresponding to mutually commuting generators.
\end{abstract}

\maketitle
\section{Introduction}\label{introduction}
One of the central objectives of quantum metrology concerns improving measurement precision with finite sized ensembles~\cite{giovannetti2011advances,Toth2014information,degen2017quantum,pezze2018quantum}. Most previous investigations have focused on single parameter estimation, of which the standard quantum limit (SQL) or the classical limit, $1/\sqrt N$, represents the minimal phase uncertainty achievable in an interferometric measurement using an ensemble of $N$ uncorrelated particles~\cite{giovannetti2006quantum}. Recently, the problem of estimating multiple parameters has attracted much interests~\cite{szczykulska2016multi,spagnolo2012quantum,humphreys2013quantum,pinel2013quantum,kim2013joint,vidrighin2014joint,crowley2014tradeoff,yao2014multiple,zhang2014quantum,yue2014quantum,
berry2015quantum,liu2015quantum,baumgratz2016quantum,gagatsos2016gaussian,ragy2016compatibility,knott2016local,ciampini2016quantum,liu2016quantum,kok2017role,yousefjani2017estimating,liu2017control,
pezze2017optimal,vrehavcek2017multiparameter,zhang2017quantum,zhuang2017entanglement,kura2018finite,zyou2018joint,liu2018loss,bradshaw2018ultimate,nichols2018multiparameter,zhuang2018multiparameter,
proctor2018multiparameter,gessner2018sensitivity,yang2018optimal,ge2018distributed,Polino19}, where the focus shifts to finding efficient strategies for estimating parameters corresponding to multiple commuting or non-commuting generators as precisely as possible. Potential applications of such studies include quantum imaging~\cite{humphreys2013quantum,yue2014quantum,vrehavcek2017multiparameter}, sensor networks~\cite{proctor2018multiparameter,ge2018distributed}, measurements of multidimensional fields~\cite{baumgratz2016quantum}, and joint measurements of multiple quadratures~\cite{kim2013joint,zyou2018joint,liu2018loss,bradshaw2018ultimate}, etc.

The main tasks in multi-parameter estimation are to generate an input quantum state capable of realizing the optimal precision limited by the laws of quantum mechanics, and to find a corresponding measurement scheme that achieves this precision. In the language of estimation theory, the former obtains a quantum state with the lowest quantum Cram\'{e}r-Rao bound (QCRB) for a set of parameters to be estimated, while the latter provides measurement results of which the Cram\'{e}r-Rao bound (CRB) equals the QCRB. For estimation of a single parameter, the latter can always be fulfilled using interferometry in which the second beam splitter acts as the inverse transformation (time reversed operation) to the first~\cite{macri2016loschmidt}, such as in an atomic Ramsey interferometer and an optical Mach-Zehnder interferometer. However, the same does not apply in general for multi-parameter estimation.

\begin{figure}
 \includegraphics[width=0.95\columnwidth]{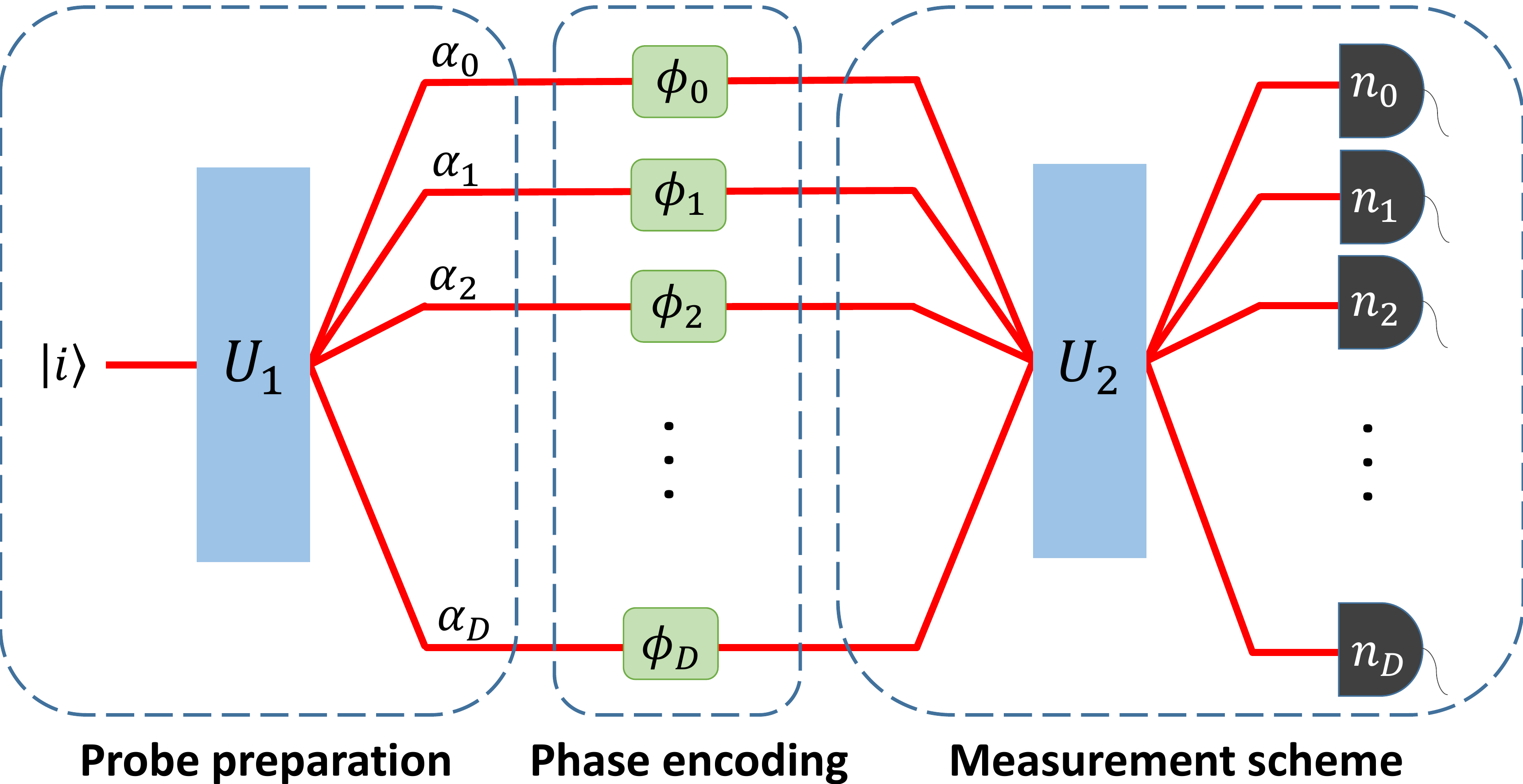}
 \caption{\label{framework} A standard (D+1)-mode interferometer for unentangled particles. The interferometer starts with a pure single mode state $\left| i \right\rangle$ followed by a unitary transformation $U_1$ (linear beam splitter), phase accumulation, and a second unitary transformation $U_2$ (combining), and ends with particle number detection in every mode.}
\end{figure}

This work considers the more specific case of multi-mode interferometry for estimating a set of parameters corresponding to mutually commuting generators using unentangled particles (as illustrated in Fig.~\ref{framework}). A probe state is generated by splitting a pure single-mode state $\left| i \right\rangle$ into multiple modes using a multi-mode beam splitter represented by a unitary transformation $U_1$. The prepared $(D+1)$-mode probe state then undergoes phase accumulation, and is subsequently transformed by $U_2$ at the second beam splitter. The measured particle number distributions at the $(D+1)$-outputs are used to estimate the $D$ parameters in the end. Unlike the case of single parameter estimation, setting $U_2=U_1^\dagger=U_1^{-1}$ does not guarantee CRB will be equal to the QCRB in general when $D>1$. Instead for a given $U_1$, $U_2$ has to be optimized numerically to reach QCRB. This becomes a cumbersome and tedious job particularly when the number of parameters to be estimated is large. 

As a main result to be reported in this paper, we prove that the Ramsey interferometric scheme can be straightforwardly generalized to estimation of multiple parameters (associated with commuting generators) using multi-mode pure states, if (but not iff) $U$ is made orthogonal, i.e. when $U$ is unitary and has only real matrix elements. We also illustrate how such orthogonal $U$ can be constructed experimentally in a simple and scalable way. The influences of particle number fluctuation and detection noise will also be discussed.

So far, most measurement schemes which saturate the QCRB for multi-parameter estimation, if they exist, are found on a case by case basis. Important progresses have been made in this direction recently~\cite{humphreys2013quantum,pezze2017optimal,zyou2018joint,zhuang2018multiparameter,yang2018optimal}. However many of the proposed measurement schemes are either not directly implementable or experimentally prohibitive, particularly when they involve measurements on entangled particles~\cite{szczykulska2016multi,spagnolo2012quantum,humphreys2013quantum,pinel2013quantum,kim2013joint,vidrighin2014joint,crowley2014tradeoff,yao2014multiple,zhang2014quantum,yue2014quantum,
berry2015quantum,liu2015quantum,baumgratz2016quantum,gagatsos2016gaussian,ragy2016compatibility,knott2016local,ciampini2016quantum,liu2016quantum,kok2017role,yousefjani2017estimating,liu2017control,
pezze2017optimal,vrehavcek2017multiparameter,zhang2017quantum,zhuang2017entanglement,kura2018finite,zyou2018joint,liu2018loss,bradshaw2018ultimate,nichols2018multiparameter,zhuang2018multiparameter,
proctor2018multiparameter,gessner2018sensitivity,yang2018optimal,ge2018distributed,Polino19}. Therefore, generalization of Ramsey interferometry to multi-parameter estimation represents an interesting and timely advance.

This article is organized as follows: Sec.~\ref{QCRB} defines the problem we consider and gives the QCRB of a multi-mode probe state. Sec.~\ref{measurement} proves that for an orthogonal $U_1$, the CRB from setting $U_2=U_1^{-1}$ is equal to the QCRB. Sec.~\ref{probe} illustrates how to determine the optimal probe state that gives the lowest QCRB. In Sec.~\ref{experiment}, we show how an orthogonal $U$ can always be constructed experimentally in a simple and scalable manner in an optical or atomic system. Finally, we consider the influence of particle number fluctuation and detection noise on the multi-mode Ramsey interferometer in Sec.~\ref{nfluctuation} and Sec.~\ref{dnoise}, respectively. The article ends with appendices~\ref{appendix_measurement} and~\ref{appendix_probe} containing further calculation details.

\section{General framework and the QCRB of a given probe state}\label{QCRB}
In this section, we define the problem we consider and give the QCRB for a given probe state. As shown in Fig.~\ref{framework}, the parameters we consider are encoded into quantum states with $D+1$ modes, which can be implemented with photons split into multiple paths, or atoms with large spins. For unentangled particles, the interferometry can be discussed in terms of self-interference of individual particles \cite{dirac1981principles}. Therefore, we consider an arbitrary single particle initial state $\left| i \right\rangle$, and a probe state $\left| {{\psi _{\rm{p}}}} \right\rangle  = {U_1}\left| i \right\rangle  = \sum\nolimits_{k = 0}^D {{\alpha _k}\left| k \right\rangle }$ after transformation $U_1$, with ${{\alpha _k}}$ being the probability amplitude in mode $k$. We assume that the probe state is pure for the time being and the interferometry is noiseless. The phase accumulation evolves the probe state into $\left| {{\psi _{\bm{\phi }}}} \right\rangle  = \sum\nolimits_{k = 0}^D {{\alpha _k}{e^{i{\phi _k}}}\left| k \right\rangle }$. Interference from the first order coherence allows $D$ (out of the $D+1$) phases to be measured in the absence of an external reference. This is often carried out by choosing an arbitrary mode, say $\left| 0 \right\rangle $, as the reference, and measuring the relative phase shifts $\theta_k\equiv\phi_k-\phi_0$ ($k=1,2,..,D$). However, if the parameters of interest ${\bm{\Theta }} \equiv \left\{ {{\Theta _1},{\Theta _2}, \cdots ,{\Theta _D}} \right\}$ are not the same as ${\bm{\theta}}\equiv\{\theta_1,\theta_2,...,\theta_D\}$, optimized precision of $\bm{\theta}$ does not necessarily give the best precision for ${\bm{\Theta }}$ in general. We assume in the following that each parameter of interest, $\Theta_k$, is a linear combination of $\left\{ {{\phi _0},{\phi _1}, \cdots ,{\phi _D}} \right\}$ in general, and the goal turns to finding a probe state that minimizes the total phase variance ${\left( {\Delta {\bm{\Theta }}} \right)^2}{\rm{ = }}\sum\limits_{k = 1}^D {{{\left( {\Delta {\Theta _k}} \right)}^2}}$.

With $\bm{\Theta}$ defined, the phases can in turn be written as $\phi_k=f_k(\bm{\Theta})$ (see Appendix~\ref{appendix_probe} for more details), and the probe state after phase accumulation becomes $\left| {{\psi _{\bm{\phi}}}} \right\rangle  = \sum\nolimits_{k = 0}^D {{\alpha _k}{e^{i  {f_k}\left( {\bm{\Theta }} \right)}}\left| k \right\rangle }$. According to multi-parameter quantum estimation theory~\cite{helstromquantum,paris2009quantum},
the lower bound of ${\left( {\Delta {\bm{\Theta }}} \right)^2}$ with an unbiased estimator is determined by the trace of the inverse of quantum Fisher information matrix (QFIM) ${{{\cal F}^Q}}$:
\begin{equation} \label{eq:QCRB}
{\left( {\Delta {{\bm{\Theta }}}} \right)^2} \ge {\rm{Tr}}\left[ {{{\left( {N_{\hat M}{{\cal F}^Q}} \right)}^{ - 1}}} \right],
\end{equation}
where $N_{\hat M}$ is the number of experiments repeated (set to 1 hereafter for simplicity). Note that, the choice of a figure of merit for precision as the trace of the inverse of ${{{\cal F}^Q}}$ in Eq.~(\ref{eq:QCRB}) is fully general, since the weight of each parameter $\phi_k$ can be adjusted by changing the coefficients of the linear combinations in $\phi_k=f_k(\bm{\Theta})$.

For a pure state $\left| {{\psi _{\bm{\phi}}}} \right\rangle$, the matrix elements of ${{{\cal F}^Q}}$ are explicitly given by~\cite{helstromquantum,paris2009quantum}
\begin{equation} \label{eq:QFIM}
{\cal F}_{l,n}^Q = 4{\rm{Re}}\left[ {\left\langle {{\partial _{{\Theta _l}}}{\psi _{\bm{\phi }}}\left| {{\partial _{{\Theta _n}}}{\psi _{\bm{\phi}}}} \right.} \right\rangle  - \left\langle {{{\partial _{{\Theta _l}}}{\psi _{\bm{\phi}}}}}
 \mathrel{\left | {\vphantom {{{\partial _{{\Theta _k}}}{\psi _{\bm{\phi}}}} {{\psi _{\bm{\Theta }}}}}}
 \right. \kern-\nulldelimiterspace}
 {{{\psi _{\bm{\phi}}}}} \right\rangle \left\langle {{{\psi _{\bm{\phi}}}}}
 \mathrel{\left | {\vphantom {{{\psi _{\bm{phi}}}} {{\partial _{{\Theta _l}}}{\psi _{\bm{\phi}}}}}}
 \right. \kern-\nulldelimiterspace}
 {{{\partial _{{\Theta _n}}}{\psi _{\bm{\phi}}}}} \right\rangle } \right],
\end{equation}
where $l,n=1,2,\cdots,D$.
The matrix elements of the $D \times D$ single-particle QFIM for $|\psi_{\bm{\phi}}\rangle$ are thus given by
\begin{widetext}
\begin{equation} \label{eq:QFIM_gen}
{\cal F}_{l,n}^Q = 4\left[ {\sum\limits_{k = 0}^D {{{\partial {f_k}\left( {\bm{\Theta }} \right)} \over {\partial {\Theta _l}}}{{\partial {f_k}\left( {\bf{\Theta }} \right)} \over {\partial {\Theta _n}}}{{\left| {{\alpha _k}} \right|}^2} - } \sum\limits_{k,k' = 0}^D {{{\partial {f_k}\left( {\bm{\Theta }} \right)} \over {\partial {\Theta _l}}}{{\partial {f_{k'}}\left( {\bm{\Theta }} \right)} \over {\partial {\Theta _n}}}{{\left| {{\alpha _k}} \right|}^2}{{\left| {{\alpha _{k'}}} \right|}^2}} } \right].
\end{equation}
\end{widetext}
The QFIM is convex and additive~\cite{gessner2018sensitivity}. For an uncorrelated but identically prepared $N$-particle product state, $\left| {{\psi _{\bm{\phi}}}} \right\rangle^{\otimes N}$, it is nothing but just the sum ($N$-times) of the single-particle QFIM. According to Eq.~(\ref{eq:QCRB}), the probe state that gives the best QCRB for estimating ${\bm{\Theta }}$ can be obtained by minimizing the trace of $\left({\cal F}^Q\right)^{-1}$, via varying $|\alpha_k|^2$ under the normalization condition $\sum\nolimits_{k = 0}^D {{{\left| {{\alpha _k}} \right|}^2}}  = 1$. It is clear from Eq.~(\ref{eq:QFIM_gen}) that the QCRB of a probe state depends only on the distribution of the particles $|\alpha_k|^2$ but not on the phase of $\alpha_k$.

\section{Ramsey interferometry for multi-parameter estimation}\label{measurement}
In the above section, we discuss how to calculate the ultimate sensitivity bound for a given beam splitting transformation $U_1$ and for a given set of parameters corresponding to mutually commuting generators. To saturate this bound, optimization of the second beam combining transformation ($U_2$) is required on a case by case basis, which is a tedious task for a large number of parameters. In this section, we prove that if (but not iff) the multi-mode unitary transformation $U_1=U$ is real (orthogonal), then $U_2 = U^\dag=U^{-1}$ followed by particle number detection afterwards gives the best precision allowed by the QCRB when $\bm{\Theta}\sim0$.

For Ramsey interferometry, the state after the full interferometric protocol (before particle number detection) is represented by $\left| {{\psi _{{\rm{out}}}}} \right\rangle  = {U^\dag }\prod\nolimits_{k = 0}^D {{e^{i\left| k \right\rangle \left\langle k \right|{f_k}\left( {\bf{\Theta }} \right)}}} U\left| i \right\rangle$. The CRB, which sets the minimal $\left( {\Delta {{\bm{\Theta }}}} \right)^2$ given a measurement scheme, can be calculated for any $U$ using the classical Fisher information matrix (CFIM)~\cite{kay1993fundamentals}
\begin{equation} \label{eq:CFIM}
{\cal F}_{l,n}^C\left( {\bm{\Theta }} \right) = \sum\limits_{m=0}^D {{1 \over {p\left( {m|{\bm{\Theta }}} \right)}}{{\partial p\left( {m|{\bm{\Theta }}} \right)} \over {\partial {\Theta _l}}}{{\partial p\left( {m|{\bm{\Theta }}} \right)} \over {\partial {\Theta _n}}}},
\end{equation}
where $p\left( {m|{\bm{\Theta }}} \right) = {\left| {\left\langle m \right|{U^\dag }\prod\limits_{k = 0}^D {{e^{i\left| k \right\rangle \left\langle k \right|{f_k}\left( {\bm{\Theta }} \right)}}} U\left| i \right\rangle } \right|^2}$ denotes the probability of finding a particle in mode $|m\rangle$ for a given $\bm{\Theta}$.

To show that a Ramsey interferometric scheme can be used to estimate multiple parameters to the SQL precision, we need to prove the CFIM given by Eq.~(\ref{eq:CFIM}) equals to QFIM given by Eq.~(\ref{eq:QCRB}) (since they correspond to the CRB and the QCRB, respectively). For small $\bm{\Theta}$, omitting the third order corrections, a Taylor series expansion around ${\bm{\Theta }} \sim 0$ gives
\begin{widetext}
\begin{equation}\label{eq:probability}
p\left( {m|{\bm{\Theta }}} \right) \simeq\begin{cases}
 \sum\limits_{k,k' = 0}^D {{f_k}\left( {\bm{\Theta }} \right){f_{k'}}\left( {\bm{\Theta }} \right)\left\langle m \right|{U^\dag }\left| k \right\rangle \left\langle k \right|U\left| i \right\rangle \left\langle i \right|{U^\dag }\left| {k'} \right\rangle \left\langle {k'} \right|U\left| m \right\rangle }&\text{$m\ne i$},\\
1 + \sum\limits_{k,k' = 0}^D {{f_k}\left( {\bm{\Theta }} \right){f_{k'}}\left( {\bm{\Theta }} \right){{\left| {\left\langle i \right|{U^\dag }\left| k \right\rangle } \right|}^2}{{\left| {\left\langle {k'} \right|U\left| i \right\rangle } \right|}^2}}  - \sum\limits_{k = 0}^D {{f_k}{{\left( {\bm{\Theta }} \right)}^2}{{\left| {\left\langle i \right|{U^\dag }\left| k \right\rangle } \right|}^2}}&
\text{$m=i$}.
\end{cases}\
\end{equation}
\end{widetext}
The derivatives of Eq.~(\ref{eq:probability}) with respect to any $\Theta_l$ can also be calculated directly (see Appendix~\ref{appendix_measurement}). Substituting Eq.~(\ref{eq:probability}) and its derivatives into the definition of the CFIM (Eq.~(\ref{eq:CFIM})) gives a complicated equation that looks vastly different from the QFIM of Eq.~(\ref{eq:QFIM}). Indeed, for an arbitrary $U$, the CFIM for an Ramsey interferometric scheme is not equal to the QFIM in most cases. We find, however, that when $\left\langle k \right|U\left| m \right\rangle = \left\langle m \right|U^\dag\left| k \right\rangle$ for all $k,m$ (meaning that all elements of $U$ are real and $U$ is orthogonal), the equation for the CFIM is simplified greatly and becomes (see Appendix~\ref{appendix_measurement})
\begin{widetext}
\begin{equation}
{\cal F}_{l,n}^C\left( {{\bf{\Theta }} \sim 0} \right) \simeq 4\sum\limits_{m \ne i} {\sum\limits_{k,k' = 0}^D {\frac{{\partial {f_k}\left( {\bf{\Theta }} \right)}}{{\partial {\Theta _l}}}\frac{{\partial {f_{k'}}\left( {\bf{\Theta }} \right)}}{{\partial {\Theta _n}}}\left\langle i \right|{U^\dag}\left| k \right\rangle \left\langle k \right|U\left| m \right\rangle \left\langle m \right|{U^\dag }\left| {k'} \right\rangle \left\langle {k'} \right|U\left| i \right\rangle } }.
\end{equation}
\end{widetext}
To further simplify the formula, we make use of the completeness of the basis $\left| i \right\rangle \left\langle i \right| + \sum\nolimits_{m \ne i} {\left| m \right\rangle \left\langle m \right|}  = {\bf{1}}$. This gives
\begin{widetext}
\begin{eqnarray} \label{eq:proof_supp_3}
{\cal F}_{l,n}^C\left( {{\bm{\Theta }} \sim 0} \right) &\simeq& 4\sum\limits_{m \ne i} {\sum\limits_{k,k' = 0}^D {\frac{{\partial {f_k}\left( {\bf{\Theta }} \right)}}{{\partial {\Theta _l}}}\frac{{\partial {f_{k'}}\left( {\bf{\Theta }} \right)}}{{\partial {\Theta _n}}}\left\langle i \right|{U^\dag }\left| k \right\rangle \left\langle k \right|U\underbrace {\left| m \right\rangle \left\langle m \right|}_{m \ne i}{U^\dag }\left| {k'} \right\rangle \left\langle {k'} \right|U\left| i \right\rangle } } \nonumber \\
&=& 4\left[ {\sum\limits_{k,k' = 0}^D {{{\partial {f_k}\left( {\bf{\Theta }} \right)} \over {\partial {\Theta _l}}}{{\partial {f_{k'}}\left( {\bf{\Theta }} \right)} \over {\partial {\Theta _n}}}\left\langle i \right|{U^\dag }\left| k \right\rangle \left\langle k \right|U\left( {{\bf{1}} - \left| i \right\rangle \left\langle i \right|} \right){U^\dag }\left| {k'} \right\rangle \left\langle {k'} \right|U\left| i \right\rangle } } \right]\nonumber \\
&=& 4\left[ {\sum\limits_{k = 0}^D {{{\partial {f_k}\left( {\bf{\Theta }} \right)} \over {\partial {\Theta _l}}}{{\partial {f_{k}}\left( {\bf{\Theta }} \right)} \over {\partial {\Theta _n}}}\left\langle i \right|{U^\dag }\left| k \right\rangle \left\langle k \right|U\left| i \right\rangle  - \sum\limits_{k,k' = 0}^D {{{\partial {f_k}\left( {\bf{\Theta }} \right)} \over {\partial {\Theta _l}}}{{\partial {f_{k'}}\left( {\bf{\Theta }} \right)} \over {\partial {\Theta _n}}}\left\langle i \right|{U^\dag }\left| k \right\rangle \left\langle k \right|U\left| i \right\rangle \left\langle i \right|{U^\dag }\left| {k'} \right\rangle \left\langle {k'} \right|U\left| i \right\rangle } } } \right]\nonumber\\
&=& 4\left[ {\sum\limits_{k = 0}^D {{{\partial {f_k}\left( {\bf{\Theta }} \right)} \over {\partial {\Theta _l}}}{{\partial {f_{k}}\left( {\bf{\Theta }} \right)} \over {\partial {\Theta _n}}}{{\left| {{\alpha _k}} \right|}^2} - } \sum\limits_{k,k' = 0}^D {{{\partial {f_k}\left( {\bf{\Theta }} \right)} \over {\partial {\Theta _l}}}{{\partial {f_{k'}}\left( {\bf{\Theta }} \right)} \over {\partial {\Theta _n}}}{{\left| {{\alpha _k}} \right|}^2}{{\left| {{\alpha _{k'}}} \right|}^2}} } \right] = {\cal F}_{l,n}^Q.
\end{eqnarray}
\end{widetext}

In Ref.~\cite{pezze2017optimal}, Pezz{\`e} et al. found the necessary and sufficient conditions (iff) for projective measurements which saturate the QCRB of a probe state. In their language, our measurement can be described by a set of projectors $\left\{ {\left| {{\Upsilon _k}} \right\rangle \left\langle {{\Upsilon _k}} \right|} \right\}$, where $\left| {{\Upsilon _k}} \right\rangle  = U\left| k \right\rangle$. In the limit ${\bf{\Theta }} \sim 0$ and given that all elements of $U$ are real, the projectors $\left\{ {\left| {{\Upsilon _k}} \right\rangle \left\langle {{\Upsilon _k}} \right|} \right\}$ indeed satisfy the required condition given by their Eq.~(7) in \cite{pezze2017optimal}.
\section{determining the optimal probe state}\label{probe}
In this section, we demonstrate how to determine the optimal probe state. As an illustration, we consider the most common choice of $\Theta_k = \theta_k \equiv \phi_k-\phi_0$. In this case, the generator of parameter $\theta _k$ is proportional to $\left| k \right\rangle \left\langle k \right|$. Computing Eq.~(\ref{eq:QFIM_gen}) and taking the trace of $\left({\cal F}^Q\right)^{-1}$ gives (after dividing by particle number $N$) (see Appendix~\ref{appendix_probe} for more details)
\begin{equation} \label{eq:variance}
{\left( {\Delta {{\bm{\theta }}_{{\rm{}}}}} \right)^2} \ge \frac{1}{N}\left[{D\over {4{{{\left| {{\alpha _0}} \right|}^2}}}}+\sum\limits_{k = 1}^D {{1 \over {4{{\left| \alpha_k \right|}^2}}}}  \right].
\end{equation}
Minimizing Eq.~(\ref{eq:variance}) under the condition $\sum\nolimits_{k = 0}^D {{{\left| {{\alpha _k}} \right|}^2}}  = 1$ gives the optimal probe state described by
\begin{subequations} \label{eq:probe}
\begin{align}
{\left| {{\alpha _0}} \right|^2} &= {{\sqrt D } \mathord{\left/
 {\vphantom {{\sqrt D } {\left( {d + \sqrt D } \right)}}} \right.
 \kern-\nulldelimiterspace} {\left( {D + \sqrt D } \right)}}, \\
{\left| {{\alpha _k}} \right|^2} &= {1 \mathord{\left/
 {\vphantom {1 {\left( {D + \sqrt D } \right)}}} \right.
 \kern-\nulldelimiterspace} {\left( {D + \sqrt D } \right)}},{\kern 1pt} {\kern 1pt} {\kern 1pt} {\kern 1pt} {\kern 1pt} {\kern 1pt} {\kern 1pt} {\kern 1pt} {\kern 1pt} \left( \mathrm{for}~{k \ne 0} \right),
\end{align}
\end{subequations}
and the QCRB of
\begin{equation} \label{eq:SNL}
{\left( {\Delta {{\bm{\theta }}_{{\rm{opt}}}}} \right)^2} = {\left( {D + \sqrt D } \right)^2}/4N.
\end{equation}
This precision can be reached in the asymptotic regime of large ${{N_{\hat M}}}$.

For comparison, we consider an individual estimation scheme which divides the $N$ particles into $D$ equal partitions,
and uses each partition for measuring one $\theta_k$ through two-mode interferometry between $|0\rangle$ and $|k\rangle$.
Since the SQL of each $\theta_k$ in this case is $1/\sqrt {N/D}$, the lowest bound for the phase variance becomes
\begin{equation}\label{eq:SNL_ind}
{\left( {\Delta {{\bm{\theta }}_{{\rm{ind}}}}} \right)^2} = {D^2}/N.
\end{equation}
For $D=1$ as in single parameter estimation, both Eqs.~(\ref{eq:SNL}) and (\ref{eq:SNL_ind}) reduce to $1/N$ as expected (i.e. the SQL). For larger $D$, the simultaneous estimation scheme(Eq.~(\ref{eq:SNL})) always outperforms the individual estimation scheme (Eq.~(\ref{eq:SNL_ind})).

We note that the results of Eqs.~(\ref{eq:probe}) and (\ref{eq:SNL}) resemble an earlier study~\cite{humphreys2013quantum}, where Humphreys et al. considered a multi-mode entangled NOON state $\left| {{\psi _{{\rm{in}}}}} \right\rangle  = {\alpha _0}\left| {N,0, \ldots ,0} \right\rangle  + {\alpha _1}\left| {0,N, \ldots ,0} \right\rangle  +  \ldots  + {\alpha _D}\left| {0,0, \ldots ,N} \right\rangle$. They found an optimal probe defined also by Eq.~(\ref{eq:probe}) and a QCRB $N$ times smaller than Eq.~(\ref{eq:SNL}), in agreement with the typical ratio between the SQL and the Heisenberg limit (HL). Their results reduce to ours when a multi-mode NOON state for $N=1$ is considered.

The value of the reference mode ${\left| {{\alpha _0}} \right|^2}$ in Eq.~(\ref{eq:probe}) is $\sqrt D $ times larger than the other modes. Because $\phi_0$ is referenced to by all $\theta_k\equiv\phi_k-\phi_0$, the measurement variance of $\phi_0$ therefore contributes $D$ times more to ${\left( {\Delta {\bm{\theta }}} \right)^2}$ than any other uncorrelated phases \{$\phi_k$\}. Such a bias results from the choice of parameters. Consequently, Eqs.~(\ref{eq:probe}) and (\ref{eq:SNL}) cannot be always optimal if the parameters of interest are different. For instance, should we consider a different set of parameters of interest, say, ${\varphi _k}\equiv{\phi_k} - {\phi_{k-1}}~(k=1,2,..,D)$, i.e., the relative phase between the neighboring modes, repeating the above same procedures gives a minimal variance of ${\left( {\Delta {{\bm{\varphi }}_{{\rm{opt}}}}} \right)^2}{\rm{ = }}{1 \over 4N}{\left[ {\sqrt 2 \left( {D - 1} \right) + 2} \right]^2}$.
If one measures \{$\theta_k$\} instead of \{$\varphi_k$\} using the probe state given by Eq.~(\ref{eq:probe}) and then derives \{$\varphi_k$\} from \{$\theta_k$\}, the resulting phase variance would be bounded by ${\left( {\Delta {\bm{\varphi }}} \right)^2}{\rm{ = }}{1 \over 4N}\left[ {{{( {1{\rm{ + }}\sqrt D } )}^2} + 2( {D - 1} )( {\sqrt D  + D} )} \right]$(for details see Appendix~\ref{appendix_probe}), a result always larger than ${\left( {\Delta {{\bm{\varphi }}_{{\rm{opt}}}}} \right)^2}$ for $D>1$.

\section{Experimental Realization of $U$}\label{experiment}
We now illustrate how a multi-mode Ramsey interferometer can be realized experimentally in a simple and scalable way. Here, the task reduces to designing an orthogonal $U$ that generates the optimal probe state. We would illustrate our scheme first for an optical interferometer and then an atomic interferometer.
\subsection{Multi-mode optical interferometer}\label{photon}
\begin{figure}[!htp]
\includegraphics[width=1\columnwidth]{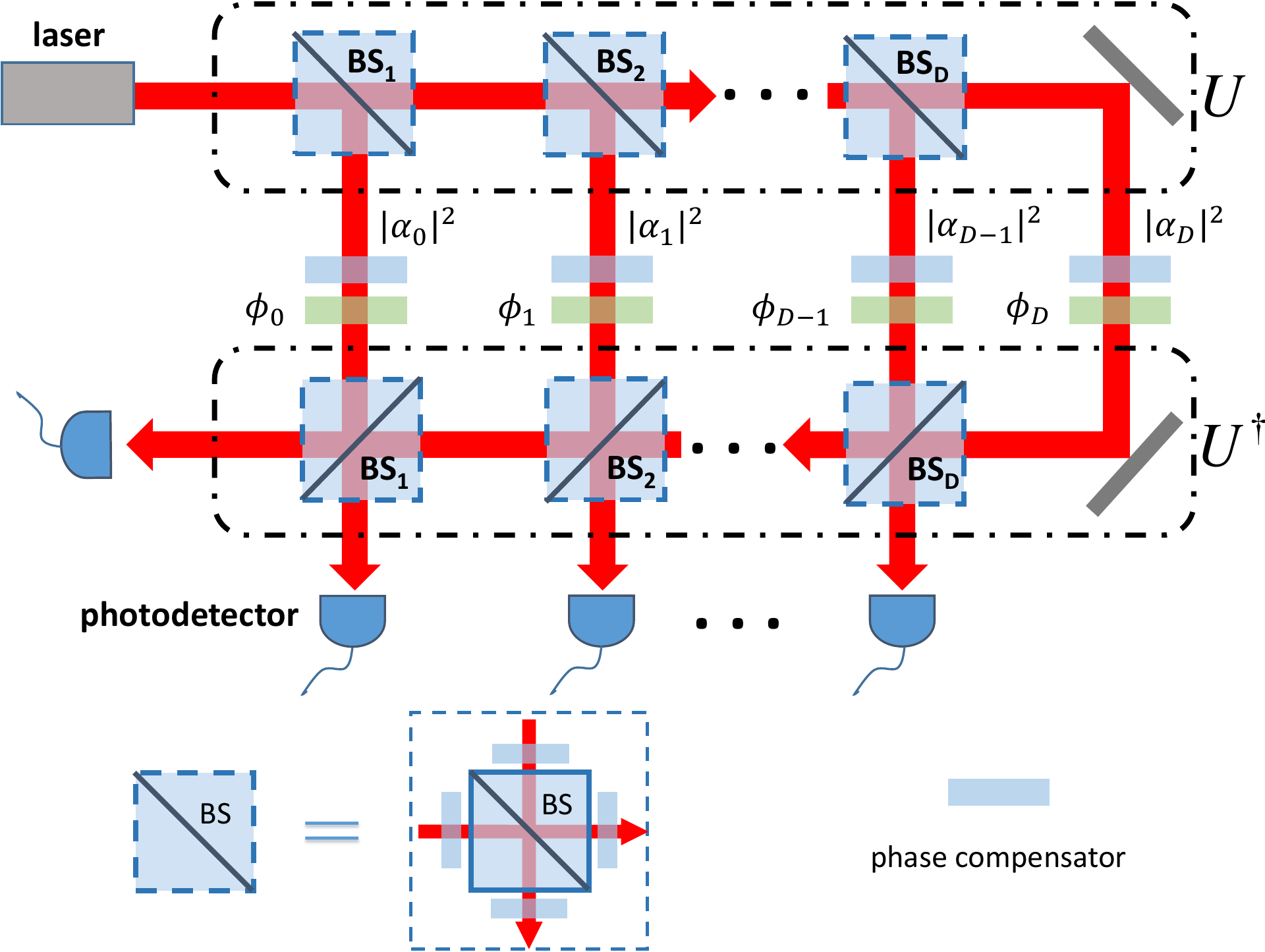}
\caption{\label{optical_interferometer} A multi-mode optical interferometer. A series of $2\times2$ beam splitters are used to form unitary transformations $U$ and $U^\dag$ with only real matrix elements. The splitting ratio of each beam splitter is chosen to distribute the input light according to the optimal probe state found. After phase accumulation, a reversed unitary transformation ${U^\dag }$ is implemented with another series of beam splitters arranged in reverse order. The interferometer ends with photocurrent detection in every output port. (Inset) Each beam splitter in dashed lines is a composite of four phase compensators and a physical beam-splitter to generate a local transformation with all real elements.}
\end{figure}

We consider a design that employs a series of $2\times2$ non-polarizing beam splitters ({$\mathrm{BS}_k$}, for $k=1,2,3,\cdots$) for splitting particles into the optimal distributions $|\alpha_k|^2$ as shown in Fig.~\ref{optical_interferometer}. In this case, each $U^{(k)}$ which represents the transformation due to beam splitter $\mathrm{BS}_k$ acts only on two of the adjacent modes, leaving other modes untouched. The overall transformation $U=U^{(D)}U^{(D-1)}\cdots U^{(1)}$ must be unitary since each lossless physical splitter $U^{(k)}$ is unitary.

To ensure that the resulting $U$ and $U^\dag$ constructed from these beam splitters are real (orthogonal), the most straightforward way is to make sure that each of the beam splitters behaves as a real $2\times2$ transformation. This criterion, which requires zero (or multiple of 2$\pi$) phase shifts for both the transmitted and reflected beams with respect to both input beams, is not automatically satisfied for any beam splitters. Fortunately, it is always possible to fulfil this criterion by adding respective phase compensating waveplate to each port of a beam splitter. After compensation, the matrix elements of a real $U^{(k)}$ become $U^{(k)}_{k,k}=U^{(k)}_{k+1,k+1}=\cos(\eta_k)$, $U^{(k)}_{k,k+1}=-U^{(k)}_{k+1,k}=\sin(\eta_k)$, and $U^{(k)}_{i,j\not\in \{k,k+1\}}=\delta_{i,j}$ (the Kronecker delta function), where ${\cos ^2}{\eta _k}$ (${\sin ^2}{\eta _k}$) represents the reflectance (transmittance) of $\mathrm{BS}_k$. Given an optimal distribution $|\alpha_k|^2$, the reflectance of $\mathrm{BS}_k$ should be chosen as
\begin{subequations}\label{eq:transformation}
\begin{align}
&|\cos {\eta _1}|^2 = {\left| {{\alpha _0}} \right|^2},\\
&|\cos {\eta _k}|^2 = \frac{{\left| {{\alpha _{k - 1}}} \right|^2}}{{1 - {{\left| {{\alpha _0}} \right|}^2} - {{\left| {{\alpha _1}} \right|}^2} \cdots  - {{\left| {{\alpha _{k - 2}}} \right|}^2} }}{\kern 1pt} {\kern 1pt} {\kern 1pt} {\kern 1pt} {\kern 1pt} {\kern 1pt} {\kern 1pt} {\kern 1pt} {\kern 1pt} {\kern 1pt} {\kern 1pt} {\kern 1pt} {\kern 1pt} \left( {k \ge 2} \right).
\end{align}
\end{subequations}

In addition, extra phase compensators are needed in every arm of the interferometer to null out the difference in optical path lengths and to tune every phase shift $\phi_k$ to the region where $ \bm{\Theta}$ can be measured most sensitively. If there is no detection noise, this region is $\bm{\Theta}$ close to zero, otherwise, it is shifted away from $\bm{\Theta}\sim0$ (see Sec.~\ref{dnoise}). We emphasize that compensating for $\phi_k$ to give the optimal sensitivity does not represent a flaw, in fact, as such tuning is needed in practically all real interferometric measurements near the SQL precision.
\subsection{Multi-mode atomic interferometer}\label{atom}
\begin{figure}
\includegraphics[width=0.95\columnwidth]{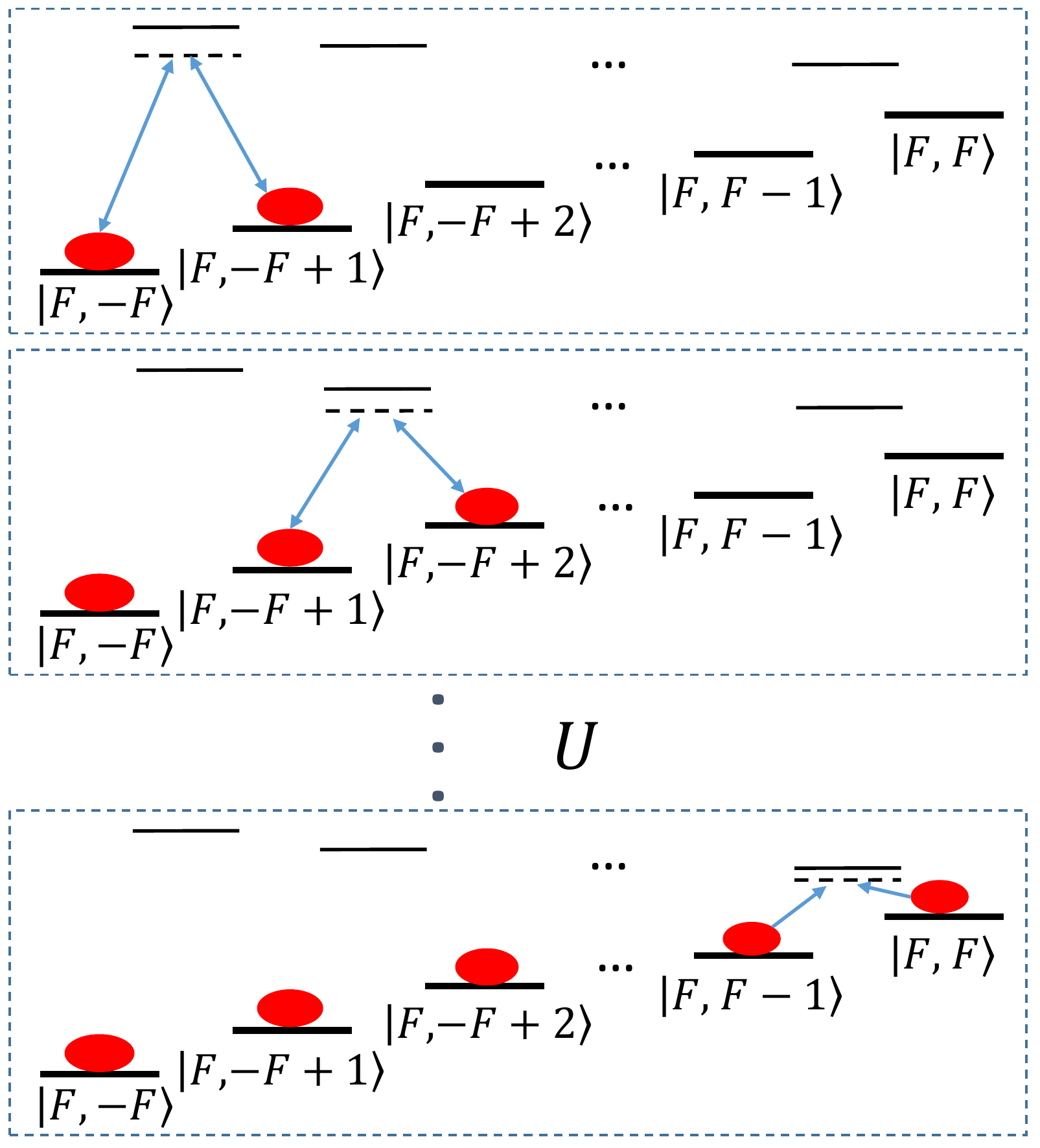}
\caption{\label{atomic_interferometer} Preparation of the optimal probe state with a sequence of two-photon Raman pulses between two adjacent states. This example starts from the state $|F,-F\rangle$, although more generally the state preparation can start from any Zeeman sublevels to reach the same final  probability amplitude distribution of the optimal state.}
\end{figure}

In interferometry of atoms with hyperfine spin $F$, $D=2F$ different parameters can be estimated. Analogous to the optical scheme, an arbitrary spin distribution can be constructed using a sequence of Rabi rotations between two adjacent Zeeman sublevels. Such rotations can be realized, for instance, using a two-photon Raman transition through an intermediate state as illustrated in Fig.~\ref{atomic_interferometer}. As long as the intermediate hyperfine levels have a different Land\'{e} g-factor from those involved in interferometry, one could perform Rabi rotations between any two adjacent sublevels by selectively detuned to a suitable intermediate states. To make sure that the individual transformation is orthogonal, every rotation should be performed along the $\sigma_y=\begin{pmatrix}0&-i\\i&0\end{pmatrix}$ direction, such that $U^{(k)}=\exp(-i\sigma_y\beta_k)=\begin{bmatrix}\cos(\beta_k)&-\sin(\beta_k)\\ \sin(\beta_k)&\cos(\beta_k)\end{bmatrix}$, within the two-level subspace. However, as each of the Zeeman sublevel exhibits different shift inside a magnetic field and thus different phase accumulation rate, one would need to keep track of the phases of every levels and to account for them when performing individual Rabi rotations. While this is possible with current technologies in cold atom experiments, the process is perhaps too cumbersome to be practical, especially when atomic spin is large.

\begin{figure}[!htp]
\includegraphics[width=0.95\columnwidth]{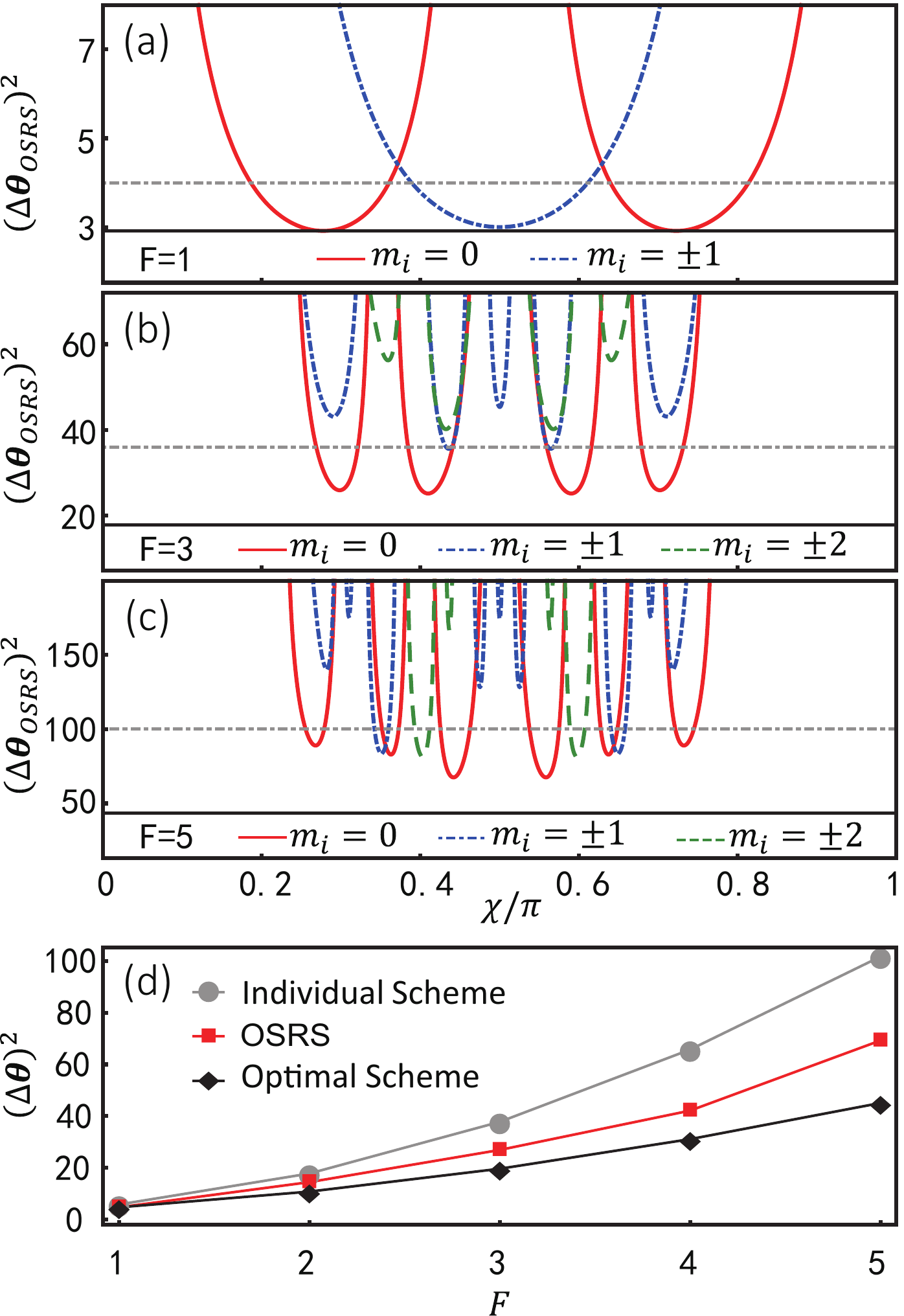}
\caption{\label{OneStep} Total measurement variance $(\Delta\bm{\theta})^2$ from OSRS for various atomic spin $F$.  (a), (b) and (c) show $(\Delta\bm{\theta})^2$ of OSRS, $U=\exp(-i F_y \chi)$ as a function of rotation angle $\chi$, for $F=1$, 3 and 5, respectively. The black solid lines and grey dash-dotted lines denotes $(\Delta\bm{\theta}_\mathrm{opt})^2$ (Eq.~(\ref{eq:SNL})) and $(\Delta\bm{\theta}_\mathrm{ind})^2$ (Eq. (\ref{eq:SNL_ind})), respectively. The legends show the corresponding initial state $|F,{m_i}\rangle$ before applying $U$. Irrespective of $m_i$, the phase shifts $\bm{\theta}$ are always defined with respect to the reference mode $|F,0\rangle$. (d) Comparison between the optimal $(\Delta\bm{\theta})^2$ from OSRS to $(\Delta\bm{\theta}_\mathrm{ind})^2$ and $(\Delta\bm{\theta}_\mathrm{opt})^2$ for $F = \left\{ {1,2,3,4,5} \right\}$. The OSRS is found to be on par with the optimal simultaneous scheme only for $F=1$, but it always performs better than the individual measurement scheme. $N=1$ for all figures.}
\end{figure}

For the aforementioned reasons, we restrict the transformation in the following to a single-pulse multi-mode Rabi rotation over an angle $\chi$ along the $F_y$ direction (since the corresponding matrix $U=\exp(-i F_y \chi)$ is always orthogonal for any atomic spin $F$), and study the performance of the Ramsey interferometric protocol for measuring $\bm{\theta}$. Experimentally, such a $F_y$ rotation can be realized using a radio-frequency resonant with adjacent Zeeman sublevels, when the quadratic Zeeman shift is negligible. It transforms the initial state $\left| {F,{m_i}} \right\rangle$ into $\left| {{\psi _{\rm{p}}}} \right\rangle  = \sum\nolimits_{m =  - F}^F {d_{{m_i},m}^F\left( \chi  \right)\left| {F,m} \right\rangle }$ with the Wigner's (small) d-matrix. According to Eq.~(\ref{eq:variance}), the QCRB of this state is given by
\begin{equation} \label{eq:QCRBOSSC}
{1 \over {4N}}\left[ {{{2F - 1} \over {|d_{{m_0},{m_i}}^F\left( \chi  \right){|^2}}} + \sum\limits_{m =  - F}^F {{1 \over {|d_{{m_0},m}^F\left( \chi  \right){|^2}}}} } \right],
\end{equation}
when $|F,m_0\rangle$ is chosen as the reference mode. Figures ~\ref{OneStep} (a), (b), and (c) present the values of Eq.~(\ref{eq:QCRBOSSC}) for
$F = 1$, $3$, and $5$, respectively (for $m_0=0$). This one-step-rotation scheme (OSRS), which employs the limited family of a single SU(2) transformation, is found to always outperform the individual measurement scheme (Eq.~(\ref{eq:SNL_ind}), grey dashed horizontal line)
using a suitable initial state $\left| {F,{m_i}} \right\rangle$ and a rotation angle $\chi$, at least up to $F=5$ (Fig.~\ref{OneStep}(d)). The same conclusion is reached for parameters $\{\varphi_{k}\}$.

Such multi-parameter estimation scheme can be useful when atoms are subjected to different sources of phase shifts simultaneously, as for example, with spin-1 $^{87}$Rb atoms dressed by near-resonant microwaves while under a static magnetic field~\cite{luo2017deterministic,zou2018beating}, or spin-9/2 $^{87}$Sr atoms placed in an optical lattice with polarization dependent light shifts, and collisions with background or non-condensed atoms.
\section{The effects of particle number fluctuation}\label{nfluctuation}
In this section, we discuss the influence of particle number fluctuation of the probe state. Since quantum states with a definite large particle number are often difficult to prepare, we consider the situation when the particle number of the probe state fluctuates. Due to the superselection rule, such input state represents nothing but an incoherent superposition of different Fock state ${\rho _{{\rm{in}}}} = {\mathop \oplus\nolimits_{N = 0}^{ + \infty }} {{Q_N}{\rho ^{\left( N \right)}}} $  in the absence of number coherences in the probe state and/or in the measurement strategy~\cite{jarzyna2012quantum,hyllus2010entanglement,pezze2015phase}, where ${{\rho ^{\left( N \right)}}}$ is the density matrix of the $N$-particle state and ${{Q_N}}$ the probability of having $N$ particles. For coherent light of photons or an atomic Bose-Einstein condensate, the particle number obeys Poisson distribution with the probability ${Q_N} = {e^{ - \bar N}}{{\bar N}^N}/N!$, where ${\bar N}$ denotes the mean particle number. Since QFIM is additive under a direct sum of density matrix ${\rho ^{\left( N \right)}}$ in orthogonal subspaces~\cite{Toth2014information}, ${\cal F}^Q(\rho _\mathrm{in})={{\cal F}^Q}\left[ {\mathop  \oplus \nolimits_N {Q_N}{\rho ^{\left( N \right)}}} \right] = \sum\nolimits_N {{Q_N}{{\cal F}^Q}\left[ {{\rho ^{\left( N \right)}}} \right]}$. For unentangled $N$-particle states of the form $\rho^{(N)}=\rho_\mathrm{single}^{\otimes N}=\left[|\psi_{\rm{p}}\rangle\langle\psi_{\rm{p}}|\right]^{\otimes N}$, ${{\cal F}^Q}\left[ {{\rho ^{\left( N \right)}}} \right]=N{\cal F}_{{\rm{single}}}^Q$, with ${\cal F}_{{\rm{single}}}^Q$ being the QFIM of the single particle probe state $\rho_{\rm single}$. One has therefore $ {\cal F}^Q(\rho _\mathrm{in}) = \sum\nolimits_N {{Q_N}N} {\cal F}_{{\rm{single}}}^Q = {\cal F}_{{\rm{single}}}^Q\bar N$. Similarly, it can be readily shown that ${{\cal F}^C} = \sum\nolimits_N {{Q_N}} {\cal F}^C[\rho^{\otimes N}_\mathrm{single}] = {\cal F}_{{\rm{single}}}^C\bar N$  for input state $\rho _{\rm{in}}$~\cite{Varenna} if particle numbers in all output ports are measured without detection noise. Since the CFIM of a single particle probe, ${\cal F}_{\rm{single}}^C \approx {\cal F}_{\rm{single}}^Q$ in our scheme,  ${\cal F}^C$ also equals approximately to ${\cal F}^Q$ for probe state with fluctuating particle number and all our conclusions outlined above remain intact.

\section{The influence of detection noise}\label{dnoise}
The conclusions in Sec.~\ref{measurement} and Appendix~\ref{appendix_measurement} are reached assuming noiseless particle number detections. When detection noise is taken into consideration, the optimal sensitivity typically shifts away from $\bm{\Theta}\sim 0$. For example, for single parameter estimation using Ramsey interferometry in an atomic clock, the measurement is usually performed near $\theta \sim \pi/2$, a region least sensitive to detection noise.

Here, we study numerically the effects of detection noise to the multi-parameter Ramsey interferometry using the example of two parameter estimation. We consider estimation of ${\theta _1}$ and ${\theta_2}$ using the optimal probe state given by Eq.~(\ref{eq:probe}). Starting from the initial state $|0\rangle=(0,1,0)^\dag$, we choose an orthogonal $U$ given by the SU(2) rotation of a spin-1 system along $F_y$-direction
\begin{equation}
U = \exp \left( { - i{F_y}\chi } \right) = \begin{pmatrix}
\frac{1}{2} + \frac{{\cos\chi}}{2} & - \frac{{\sin \chi }}{{\sqrt 2 }} & \frac{1}{2} - \frac{{\cos\chi}}{2}\\
\frac{{\sin \chi }}{{\sqrt 2 }} & \cos \chi  & - \frac{{\sin \chi }}{{\sqrt 2 }}\\
\frac{1}{2} - \frac{{\cos\chi}}{2} &  \frac{{\sin \chi }}{{\sqrt 2 }} & \frac{1}{2} + \frac{{\cos\chi}}{2} \\
\end{pmatrix}.
\end{equation}
Here, $\chi$ is set to $0.2774\pi$ to give the optimal probe state. The simulated procedure consists of applying $U$, phase accumulation $\exp \left[ {i\left( {{\theta _1}|1\rangle \langle 1| + {\theta _2}| - 1\rangle \langle  - 1|} \right)} \right]$ ($|1\rangle  = {\left( {1,0,0} \right)^\dag }$,$| - 1\rangle  = {\left( {0,0,1} \right)^\dag }$), and $U^\dag$, followed by population detection with or without including noise.

\begin{figure}[!htp]
\includegraphics[width=1\columnwidth]{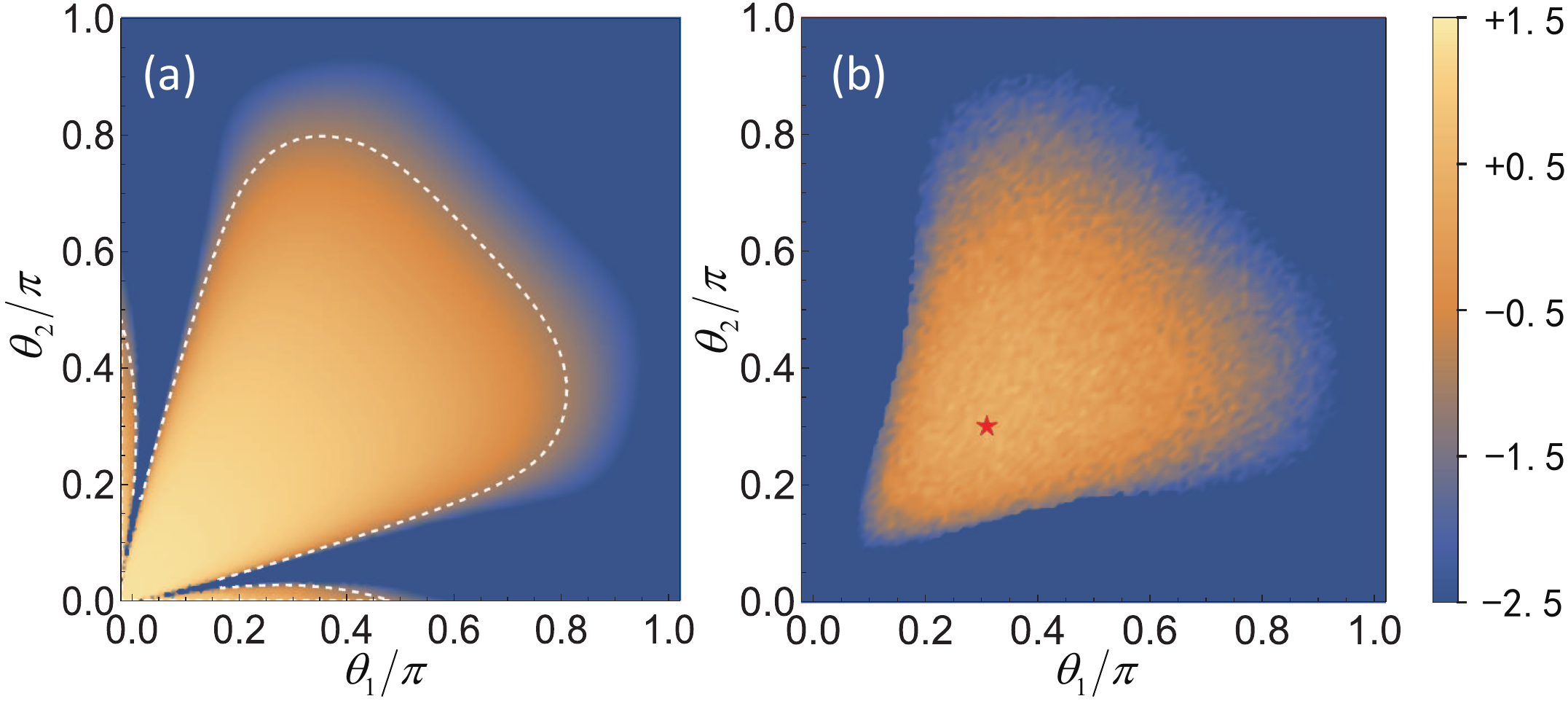}
\caption{\label{noise} Effectiveness of the proposed scheme for estimating two arbitrary $\theta_1$ and $\theta_2$ with and without detection noise. The colored figures show $\zeta$, the ratio of the CRB for the proposed scheme to the QCRB of the individual measurement scheme in negative decibels, considering (a) ideal atom-number detection and (b) atom-number resolution of $\pm14$ atoms. The area surrounded by the white dashed curve in (a) denotes the parameter space where the proposed scheme outperforms the individual measurement scheme. (b) The results of Monte Carlo simulations with $10^4$ atoms and 1000 simulated experimental runs. The star denotes the position where the minimum ${\left(\Delta\bm{\theta }\right)^2}$ occurs, which is no longer at $\bm{\theta}\sim 0$.}
\end{figure}

When there exists no detection noise, the CFIM (Eq.~(\ref{eq:CFIM})) of the aforementioned protocol is directly computed and the trace of its inverse is used to obtain the CRB of ${\left( {\Delta {\bm{\theta }}} \right)^2}$. Figure~\ref{noise}(a) compares the value of the corresponding result to the QCRB of the individual measurement scheme (Eq.~(\ref{eq:SNL_ind})) for $\{\theta_1$,$\theta_2\}\in (0,\pi)$, illustrated by the parameter $\zeta =-10\log_{10}\left[(\Delta\bm{\theta})^2/(\Delta\bm{\theta}_{\rm{ind}})^2 \right]$. The region surrounded by the white dashed curve represents the $\{\theta_1$,$\theta_2\}$-space where the proposed scheme outperforms the individual estimation scheme. It shows that the proposed scheme works well even for ${\bm{\theta}}$ far away from zero. We emphasize that the probe state defined by Eq.~(\ref{eq:probe}) gives always the best QCRB for any ${\bm{\theta}}$. However, application of the reversed transformation $U^\dag$ followed by a population measurement is not necessary the optimal measurement scheme when ${\bm{\theta}}$ is away from zero, which explains the deficiency of the scheme over some parameter space.

When detection noise is present, we numerically simulate the estimation process of the two parameters $\{\theta_1,\theta_2\}$ using $10^{4}$ three-mode (spin-$1$) atoms with a detection resolution (noise) of $14$ atoms (typical numbers achievable in cold-atom experiments \cite{luo2017deterministic,zou2018beating}). For each pair of $\{\theta_1,\theta_2\}$, we first compute the probability of detecting an atom in the output mode $m$, $p\left( {m|{\theta _1},{\theta _2}} \right) = {\left| {\left\langle m \right|{U^\dag }\exp \left[ {i\left( {{\theta _1}\left| 1 \right\rangle \left\langle 1 \right| + {\theta _2}\left| { - 1} \right\rangle \left\langle { - 1} \right|} \right)} \right]U\left| 0 \right\rangle } \right|^2}$. For each run, we perform Monte-Carlo simulation on the outcome for each of the $10^4$ atoms according to the distribution of $p\left( {m|{\theta _1},{\theta _2}} \right)$ and obtain $N_m^\prime$ (the total number of particles in mode $m$ without detection noise). We then add to $N_m^\prime$ a random detection noise featuring a normal distribution with an average of zero and a standard deviation of 14 to obtain $N_m$. The maximal likelihood method (which can saturate the CRB~\cite{kay1993fundamentals} in the asymptotic limit and is unbiased) is then used to estimate $\{\theta_1,\theta_2\}$. The likelihood function given by $L\left( {{\theta _1},{\theta _2}|{N_1},{N_0},{N_{ - 1}}} \right){\rm{ = }}\prod\nolimits_{m = 1,0, - 1} {p{{\left( {m|{\theta _1},{\theta _2}} \right)}^{{N_m}}}}$ is maximized by varying $\{\theta_1,\theta_2\}$ to obtain the estimated $\{\theta_1^\prime,\theta_2^\prime\}$. The simulation is repeated independently over $N_{\hat M} = 1000$ times. The estimated results from the 1000 simulations are then used to obtain ${\left( {\Delta {\bm{\theta }}} \right)^2}$, whose ratio to the QCRB of the individual measurement scheme using the same number of particles, is shown in Fig.~\ref{noise}(b). Although detection noise degrades the sensitivity of the proposed scheme, the discussed scheme is seen to maintain its advantage over the individual measurement scheme over a large parameter space. Similar to typical single parameter estimation scenario, the position of the minimum ${\left( {\Delta {\bm{\theta }}} \right)^2}$ is seen to shift away from zero. The star in Fig.~\ref{noise}(b) (near ${\theta _1} = {\theta _2} = 0.3\pi$) denotes the position of the maximum precision for the scenario we consider, where it is $\zeta \simeq 0.6{\rm{dB}}$ more sensitive than the individual measurement scheme, but is 0.77\ dB less than the optimal $\zeta\simeq1.37{\rm{dB}}$ for noiseless detection.

In short, the influence of detection noise to a multi-mode Ramsey interferometer is similar to  that to a single-mode Ramsey interferometer.

\section{Summary}\label{summary}
In summary, we show that the Ramsey interferometric scheme can be extended to estimation of multiple parameters (associated with commuting generators) using multi-mode pure states, if (but not iff) the multi-mode beam splitter $U$ is orthogonal, i.e. all matrix elements of $U$ are real and $UU^\dag=\bf{1}$. We then discuss how to obtain the optimal probe state, and how to construct $U$ experimentally in a simple and scalable manner. We find that the proposed scheme remains intact even under particle number fluctuation and detection noise. The results of this study can be useful to applications in multi-mode optical sensing and quantum phase imaging.

This work is supported by the National Key R\&D Program of China (Grant No. 2018YFA0306503 and No. 2018YFA0306504)
and the NSFC (Grant No. 91636213, No. 91736311, No. 11574177, No. 91836302, and No. 11654001).


\begin{widetext}

\onecolumngrid
\appendix
\renewcommand{\thesubsection}{\arabic{subsection}}

\section{Multi-mode Ramsey interferometric measurement scheme}\label{appendix_measurement}
\def\theequation{A\arabic{equation}}
\setcounter{equation}{0}
In this section, we show in detail that the proposed Ramsey-like multi-mode interferometric scheme with particle number measurement can always saturate the QCRB for small phase shift ${\bf{\Theta }}$, given that the matrix elements of the beam-splitting unitary transformation $U$ are real (or $U$ is orthogonal). The proposed scheme starts with splitting an initial state $\left| i \right\rangle$ by a unitary transformation $U$, followed by a phase accumulation process and a reversed transformation $U^\dag$, and finally ends with measuring the projection probability in mode $|m\rangle$. The projection probability in mode $|m\rangle$ after the ramsey interferometer can be explicitly written as
\begin{equation} \label{eq:pro_supp}
p\left( {m|{\bm{\Theta }}} \right) = {\left| {\left\langle m \right|{U^\dag }\prod\limits_{k = 0}^D {{e^{i\left| k \right\rangle \left\langle k \right|{f_k}\left( {\bm{\Theta }} \right)}}} U\left| i \right\rangle } \right|^2}.
\end{equation}
Omitting the third order corrections, a Taylor series expansion around ${\bm{\Theta }} \sim 0$ gives for $m\ne i$
\begin{equation} \label{eq:taylor1_supp}
p\left( {m|{\bm{\Theta }}} \right) \simeq \sum\limits_{k,k' = 0}^D {{f_k}\left( {\bm{\Theta }} \right){f_{k'}}\left( {\bm{\Theta }} \right)\left\langle m \right|{U^\dag }\left| k \right\rangle \left\langle k \right|U\left| i \right\rangle \left\langle i \right|{U^\dag }\left| {k'} \right\rangle \left\langle {k'} \right|U\left| m \right\rangle },
\end{equation}
and for $m=i$
\begin{equation} \label{eq:taylor2_supp}
p\left( {i|{\bm{\Theta }}} \right) \simeq 1 + \sum\limits_{k,k' = 0}^D {{f_k}\left( {\bm{\Theta }} \right){f_{k'}}\left( {\bm{\Theta }} \right){{\left| {\left\langle i \right|{U^\dag }\left| k \right\rangle } \right|}^2}{{\left| {\left\langle {k'} \right|U\left| i \right\rangle } \right|}^2}}  - \sum\limits_{k = 0}^D {{f_k}{{\left( {\bm{\Theta }} \right)}^2}{{\left| {\left\langle i \right|{U^\dag }\left| k \right\rangle } \right|}^2}}.
\end{equation}
The derivatives $p\left( {i|{\bm{\Theta }}} \right)$ with respect to any ${{\Theta _l}}$ is given by
\begin{equation} \label{eq:taylor_derivative_supp_0}
{{\partial p\left( {i|{\bm{\Theta }}} \right)} \over {\partial {\Theta _l}}} = 2\left[\sum\limits_{k,k' = 0}^D {{{\partial {f_k}\left( {\bm{\Theta }} \right)} \over {\partial {\Theta _l}}}{f_{k'}}\left( {\bm{\Theta }} \right) |\left\langle i \right|{U^\dag }\left| k \right\rangle |^2 \left\langle i \right|{U^\dag }\left| {k'} \right\rangle|^2 - \sum\limits_{k = 0}^D {\partial {f_k}\left( {\bm{\Theta }} \right) \over {\partial {\Theta _l}}} {{f_k}{{\left( {\bm{\Theta }} \right)}}{{\left| {\left\langle i \right|{U^\dag }\left| k \right\rangle } \right|}^2}}}\right],
\end{equation}
and the derivatives with respect to any ${{\Theta _l}}$ for $m \ne i$ can be calculated as
\begin{eqnarray} \label{eq:taylor_derivative_supp_1}
{{\partial p\left( {m|{\bm{\Theta }}} \right)} \over {\partial {\Theta _l}}} &=& \sum\limits_{k,k' = 0}^D {{{\partial {f_k}\left( {\bm{\Theta }} \right)} \over {\partial {\Theta _l}}}{f_{k'}}\left( {\bm{\Theta }} \right)\left\langle m \right|{U^\dag }\left| k \right\rangle \left\langle k \right|U\left| i \right\rangle \left\langle i \right|{U^\dag }\left| {k'} \right\rangle \left\langle {k'} \right|U\left| m \right\rangle } \nonumber \\
&+&  \sum\limits_{k,k' = 0}^D {{{\partial {f_{k'}}\left( {\bf{\Theta }} \right)} \over {\partial {\Theta _l}}}{f_k}\left( {\bm{\Theta }} \right)\left\langle m \right|{U^\dag }\left| k \right\rangle \left\langle k \right|U\left| i \right\rangle \left\langle i \right|{U^\dag }\left| {k'} \right\rangle \left\langle {k'} \right|U\left| m \right\rangle }.
\end{eqnarray}
The two terms in Eq.~(\ref{eq:taylor_derivative_supp_1}) are equivalent if all the matrix elements of $U$ (${\left\langle k \right|U\left| m \right\rangle }$) are real numbers. In this case, Eq.~(\ref{eq:taylor_derivative_supp_1}) can be simplified by summing up two terms as
\begin{equation} \label{eq:taylor_derivative_supp}
\frac{{\partial p\left( {m|{\bf{\Theta }}} \right)}}{{\partial {\Theta _l}}} = 2\sum\limits_{k,k' = 0}^D {\frac{{\partial {f_k}\left( {\bf{\Theta }} \right)}}{{\partial {\Theta _l}}}{f_{k'}}\left( {\bf{\Theta }} \right)\left\langle m \right|{U^\dag }\left| k \right\rangle \left\langle k \right|U\left| i \right\rangle \left\langle i \right|{U^\dag }\left| {k'} \right\rangle \left\langle {k'} \right|U\left| m \right\rangle }, {\kern 1pt} {\kern 1pt} {\kern 1pt} {\kern 1pt} {\kern 1pt} {\kern 1pt} {\kern 1pt} {\kern 1pt} {\kern 1pt} {\kern 1pt} {\kern 1pt} {\kern 1pt} {\kern 1pt} {\kern 1pt} {\kern 1pt} {\kern 1pt} {\kern 1pt} {\kern 1pt} m \ne i.
\end{equation}

By substituting Eqs.~(\ref{eq:taylor1_supp}), (\ref{eq:taylor2_supp}) and (\ref{eq:taylor_derivative_supp}) into the classical Fisher information matrix (CFIM)~\cite{kay1993fundamentals},
\begin{equation} \label{eq:CFIM_supp}
{\cal F}_{l,n}^C\left( {\bm{\Theta }} \right) = \sum\limits_m {{1 \over {p\left( {m|{\bm{\Theta }}} \right)}}{{\partial p\left( {m|{\bm{\Theta }}} \right)} \over {\partial {\Theta _l}}}{{\partial p\left( {m|{\bm{\Theta }}} \right)} \over {\partial {\Theta _n}}}},
\end{equation}
we obtain (when $\bm{\Theta}\sim0$ and ${\mathop{\rm Im}\nolimits} \left[ {\langle k|U\left| m \right\rangle } \right] = 0$)
\begin{small}
\begin{equation} \label{eq:proof_supp}
\sum\limits_{m \ne i} {\left\{ {\frac{{4\left[ {\sum\limits_{k,k',k'',k''' = 0}^D {\frac{{\partial {f_k}\left( {\bf{\Theta }} \right)}}{{\partial {\Theta _l}}}\frac{{\partial {f_{k'}}\left( {\bf{\Theta }} \right)}}{{\partial {\Theta _l}}}{f_{k''}}\left( {\bf{\Theta }} \right){f_{k'''}}\left( {\bf{\Theta }} \right)\left\langle m \right|{U^\dag }\left| k \right\rangle \left\langle k \right|U\left| i \right\rangle \left\langle i \right|{U^\dag }\left| {k''} \right\rangle \left\langle {k''} \right|U\left| m \right\rangle \left\langle m \right|{U^\dag }\left| {k'} \right\rangle \left\langle {k'} \right|U\left| i \right\rangle \left\langle i \right|{U^\dag }\left| {k'''} \right\rangle \left\langle {k'''} \right|U\left| m \right\rangle } } \right]}}{{\sum\limits_{k'',k''' = 0}^D {{f_{k''}}\left( {\bf{\Theta }} \right){f_{k'''}}\left( {\bf{\Theta }} \right)\left\langle m \right|{U^\dag }\left| {k''} \right\rangle \left\langle {k''} \right|U\left| i \right\rangle \left\langle i \right|{U^\dag }\left| {k'''} \right\rangle \left\langle {k'''} \right|U\left| m \right\rangle } }}} \right\}}.
\end{equation}
\end{small}
The term $m=i$ is missing from Eq.~(\ref{eq:proof_supp}) because ${{\partial p\left( {i|{\bm{\Theta }}} \right)} \over {\partial {\Theta _l}}} \propto \Theta^2\sim0$  while $p\left( {i|{\bm{\Theta }}} \right)\approx 1$. The numerator of Eq.~(\ref{eq:proof_supp}) can be factorized as
\begin{small}
\begin{equation} \label{eq:proof_supp_x}
{4\left[ {\sum\limits_{k,k' = 0}^D {\frac{{\partial {f_k}\left( {\bf{\Theta }} \right)}}{{\partial {\Theta _l}}}\frac{{\partial {f_{k'}}\left( {\bf{\Theta }} \right)}}{{\partial {\Theta _n}}}\langle m|{U^\dag }\left| k \right\rangle \langle k|U\left| i \right\rangle \langle m|{U^\dag }\left| {k'} \right\rangle \langle k'|U\left| i \right\rangle } \sum\limits_{k'',k''' = 0}^D {{f_{k''}}\left( {\bf{\Theta }} \right){f_{k'''}}\left( {\bf{\Theta }} \right)\langle i|{U^\dag }\left| {k''} \right\rangle \langle k''|U\left| m \right\rangle \langle i|{U^\dag }\left| {k'''} \right\rangle \langle k'''|U\left| m \right\rangle } } \right]},
\end{equation}
\end{small}
where the second summation cancels the denominator of Eq.~(\ref{eq:proof_supp}), giving a CFIM of the form
\begin{equation} \label{eq:proof_supp_2}
{\cal F}_{l,n}^C\left( {{\bf{\Theta }} \sim 0} \right) \simeq 4\sum\limits_{m \ne i} {\sum\limits_{k,k' = 0}^D {\frac{{\partial {f_k}\left( {\bf{\Theta }} \right)}}{{\partial {\Theta _l}}}\frac{{\partial {f_{k'}}\left( {\bf{\Theta }} \right)}}{{\partial {\Theta _n}}}\left\langle m \right|{U^\dag }\left| k \right\rangle \left\langle k \right|U\left| i \right\rangle \left\langle m \right|{U^\dag }\left| {k'} \right\rangle \left\langle {k'} \right|U\left| i \right\rangle } }.
\end{equation}
To further simplify the formula, we use again the condition that the matrix elements of $U$ (${\left\langle k \right|U\left| m \right\rangle }$) are real numbers, thus $\left\langle m \right|{U^\dag }\left| k \right\rangle  = \left\langle k \right|U\left| m \right\rangle$ and $\left\langle k \right|U\left| i \right\rangle  = \left\langle i \right|{U^\dag }\left| k \right\rangle$. This gives
\begin{eqnarray} \label{eq:proof_supp_3}
{\cal F}_{l,n}^C\left( {{\bm{\Theta }} \sim 0} \right) &\simeq& 4\sum\limits_{m \ne i} {\sum\limits_{k,k' = 0}^D {\frac{{\partial {f_k}\left( {\bf{\Theta }} \right)}}{{\partial {\Theta _l}}}\frac{{\partial {f_{k'}}\left( {\bf{\Theta }} \right)}}{{\partial {\Theta _n}}}\left\langle i \right|{U^\dag }\left| k \right\rangle \left\langle k \right|U\underbrace {\left| m \right\rangle \left\langle m \right|}_{m \ne i}{U^\dag }\left| {k'} \right\rangle \left\langle {k'} \right|U\left| i \right\rangle } } \nonumber \\
&=& 4\left[ {\sum\limits_{k,k' = 0}^D {{{\partial {f_k}\left( {\bf{\Theta }} \right)} \over {\partial {\Theta _l}}}{{\partial {f_{k'}}\left( {\bf{\Theta }} \right)} \over {\partial {\Theta _n}}}\left\langle i \right|{U^\dag }\left| k \right\rangle \left\langle k \right|U\left( {{\bf{1}} - \left| i \right\rangle \left\langle i \right|} \right){U^\dag }\left| {k'} \right\rangle \left\langle {k'} \right|U\left| i \right\rangle } } \right]\nonumber \\
&=& 4\left[ {\sum\limits_{k = 0}^D {{{\partial {f_k}\left( {\bf{\Theta }} \right)} \over {\partial {\Theta _l}}}{{\partial {f_{k}}\left( {\bf{\Theta }} \right)} \over {\partial {\Theta _n}}}\left\langle i \right|{U^\dag }\left| k \right\rangle \left\langle k \right|U\left| i \right\rangle  - \sum\limits_{k,k' = 0}^D {{{\partial {f_k}\left( {\bf{\Theta }} \right)} \over {\partial {\Theta _l}}}{{\partial {f_{k'}}\left( {\bf{\Theta }} \right)} \over {\partial {\Theta _n}}}\left\langle i \right|{U^\dag }\left| k \right\rangle \left\langle k \right|U\left| i \right\rangle \left\langle i \right|{U^\dag }\left| {k'} \right\rangle \left\langle {k'} \right|U\left| i \right\rangle } } } \right]\nonumber\\
&=& 4\left[ {\sum\limits_{k = 0}^D {{{\partial {f_k}\left( {\bf{\Theta }} \right)} \over {\partial {\Theta _l}}}{{\partial {f_{k}}\left( {\bf{\Theta }} \right)} \over {\partial {\Theta _n}}}{{\left| {{\alpha _k}} \right|}^2} - } \sum\limits_{k,k' = 0}^D {{{\partial {f_k}\left( {\bf{\Theta }} \right)} \over {\partial {\Theta _l}}}{{\partial {f_{k'}}\left( {\bf{\Theta }} \right)} \over {\partial {\Theta _n}}}{{\left| {{\alpha _k}} \right|}^2}{{\left| {{\alpha _{k'}}} \right|}^2}} } \right] = {\cal F}_{l,n}^Q.
\end{eqnarray}

In the first line of Eq.~(\ref{eq:proof_supp_3}), the completeness of the basis $\left| i \right\rangle \left\langle i \right| + \sum\nolimits_{m \ne i} {\left| m \right\rangle \left\langle m \right|}  = {\bf{1}}$ is invoked. The final result is identical to ${\cal F}_{l,n}^Q$ given by Eq.~(\ref{eq:QFIM_gen}). Thus this proves that the multi-mode Ramsey interferometer we consider here can always saturate the QCRB.

The above proof also explains why we limit the beam splitter to orthogonal matrix. This is crucial for the steps from Eq.~(\ref{eq:taylor_derivative_supp_1}) to Eq.~(\ref{eq:taylor_derivative_supp}) and from Eq.~(\ref{eq:proof_supp_2}) to Eq.~(\ref{eq:proof_supp_3}). Our proof by no means excludes the existence of $U$ with non-real matrix elements which saturates the QCRB. But the general structures of such $U$ are beyond our current knowledge.

In addition, the above proof also requires all elements $\left\langle m \right|{U}\left| k \right\rangle$ to be real even when $m,k\ne i$, which is the reason why it is insufficient to only require that the transformations from the input light in Fig.~\ref{optical_interferometer} to the outputs of all $BS_k$ are real. Instead, all beam splitters involved must act as real $2\times2$ transformations.

\section{The optimal probe state and the corresponding QCRB}\label{appendix_probe}
\def\theequation{B\arabic{equation}}
\setcounter{equation}{0}

As discussed in the main text, the probe state after phase accumulation takes the form $\left| {{\psi _{\bm{\phi }}}} \right\rangle  = \sum\nolimits_{k = 0}^D {{\alpha _k}{e^{i{\phi _k}}}\left| k \right\rangle }$. If the parameters of interest $\bm{\Theta}\equiv\left\{ {{\Theta _1},{\Theta _2}, \cdots ,{\Theta _D}} \right\}$ are linear combinations of $\phi _k$, the probe state can be expressed as $\left| {{\psi _{\bm{\phi }}}} \right\rangle  = \sum\nolimits_{k = 0}^D {{\alpha _k}{e^{i \cdot {f_k}\left( {\bf{\Theta }} \right)}}\left| k \right\rangle }$, where $f_k(\bm{\Theta})$ are linear functions of $\bm{\Theta}$. The derivative of the state above w.r.t. $\Theta_l$ is
\begin{equation} \label{eq:deriv_supp}
\left| {{\partial _{{\Theta _l}}}{\psi _{\bm{\phi }}}} \right\rangle  = i\sum\limits_{k = 0}^D {{{\partial {f_k}\left( {\bf{\Theta }} \right)} \over {\partial {\Theta _l}}}{\alpha _k}{e^{i \cdot {f_k}\left( {\bf{\Theta }} \right)}}\left| k \right\rangle }.
\end{equation}
For a pure state $\left| {{\psi _{\bm{\phi}}}} \right\rangle$,
the matrix elements of ${{{\cal F}^Q}}$ are explicitly given by~\cite{helstromquantum,paris2009quantum}
\begin{equation} \label{eq:QFIM_supp}
{\cal F}_{l,n}^Q = 4{\rm{Re}}\left[ {\left\langle {{\partial _{{\Theta _l}}}{\psi _{\bm{\phi }}}\left| {{\partial _{{\Theta _n}}}{\psi _{\bm{\phi}}}} \right.} \right\rangle  - \left\langle {{{\partial _{{\Theta _l}}}{\psi _{\bm{\phi}}}}}
 \mathrel{\left | {\vphantom {{{\partial _{{\Theta _k}}}{\psi _{\bm{\phi}}}} {{\psi _{\bm{\Theta }}}}}}
 \right. \kern-\nulldelimiterspace}
 {{{\psi _{\bm{\phi}}}}} \right\rangle \left\langle {{{\psi _{\bm{\phi}}}}}
 \mathrel{\left | {\vphantom {{{\psi _{\bm{phi}}}} {{\partial _{{\Theta _n}}}{\psi _{\bm{\phi}}}}}}
 \right. \kern-\nulldelimiterspace}
 {{{\partial _{{\Theta _n}}}{\psi _{\bm{\phi}}}}} \right\rangle } \right],
\end{equation}
where $l,n=1,2,\cdots,$ and $D$.
Substituting Eq.~(\ref{eq:deriv_supp}) into Eq.~(\ref{eq:QFIM_supp}) gives the matrix elements of quantum Fisher information matrix (QFIM) of $|\psi_{\bm{\phi}}\rangle$

\begin{equation} \label{eq:QFIM_gen_supp}
{\cal F}_{l,n}^Q = 4\left[ {\sum\limits_{k = 0}^D {{{\partial {f_k}\left( {\bm{\Theta }} \right)} \over {\partial {\Theta _l}}}{{\partial {f_k}\left( {\bf{\Theta }} \right)} \over {\partial {\Theta _n}}}{{\left| {{\alpha _k}} \right|}^2} - } \sum\limits_{k,k' = 0}^D {{{\partial {f_k}\left( {\bm{\Theta }} \right)} \over {\partial {\Theta _l}}}{{\partial {f_{k'}}\left( {\bm{\Theta }} \right)} \over {\partial {\Theta _n}}}{{\left| {{\alpha _k}} \right|}^2}{{\left| {{\alpha _{k'}}} \right|}^2}} } \right].
\end{equation}

In the case of $\Theta_k = \theta_k = \phi_k-\phi_0$ $(k=1,2,\cdots,D)$, one can choose $f_0(\bm{\Theta})=\phi_0, f_k(\bm{\Theta})=\Theta_k+\phi_0$ for $k=1,2,\cdot\cdot\cdot,D$. The matrix elements of $N$-particle QFIM can be calculated with ${{\partial {f_k}\left( {\bf{\Theta }} \right)} \over {\partial {\Theta _l}}}{\rm{ = }}{\delta _{k,l}}$, leading to the result
\begin{equation} \label{eq:element_supp}
{\cal F}_{n,l}^Q = 4N\left[ {{{\left| {{\alpha _l}} \right|}^2}{\delta _{l,n}} - {{\left| {{\alpha _l}} \right|}^2}{{\left| {{\alpha _n}} \right|}^2}} \right].
\end{equation}
Note that since there are $D+1$ $\phi_k$ but only $D$ $\Theta_k$ to be estimated, one of the $f_k(\bm{\Theta})$ can chosen at will without affecting the final results. For example, for $\Theta_k = \phi_k-\phi_0$, one can also choose $f_1(\bm{\Theta})=\phi_1, f_0(\bm{\Theta})=\phi_1-\Theta_1$, $f_k(\bm{\Theta})=\Theta_k+\Theta_1-\phi_1$ for $k=2,\cdot\cdot\cdot,D$. Substituting the so-chosen $f_k(\bm{\Theta})$ into Eq.~(\ref{eq:QFIM_gen_supp}) gives the same results as Eq.~(\ref{eq:element_supp}).

The inverse of Eq.~(\ref{eq:element_supp}) can be obtained analytically as
\begin{equation} \label{eq:inv_QFIM_supp}
\frac{1}{N}\left[ {\rm{diag}}\left( {{1 \over {4{{\left| {{\alpha _1}} \right|}^2}}},{1 \over {4{{\left| {{\alpha _2}}
\right|}^2}}} \cdots {1 \over {4{{\left| {{\alpha _D}} \right|}^2}}}} \right){\rm{ + }}{{{\bf{G}}} \over {{{4\left| {{\alpha _0}} \right|}^2}}}\right],
\end{equation}
where $\bf{G}$ is a $d \times d$ all-ones matrix. Taking the trace of Eq.~(\ref{eq:inv_QFIM_supp}) gives
\begin{equation} \label{eq:variance_supp}
{\left( {\Delta {{\bm{\theta }}_{{\rm{}}}}} \right)^2} \ge \frac{1}{N}\left[{D\over {4{{{\left| {{\alpha _0}} \right|}^2}}}}+\sum\limits_{k = 1}^D {{1 \over {4{{\left| \alpha_k \right|}^2}}}}  \right].
\end{equation}
To find the optimal probe state and the corresponding total phase variance (optimal QCRB), we minimize Eq.~(\ref{eq:variance_supp}) under the normalization condition $\sum\nolimits_{k = 0}^D {{{\left| {{\alpha _k}} \right|}^2}} = 1$. Setting the derivatives ${\partial \left[ {{{\left( {\Delta {\bm{\theta }}} \right)}^2}} \right]/\partial {{\left| {{\alpha _k}} \right|}^2}}$ to zero for any $k=1,2,\cdots,$ and $D$ gives a set of equations,
\begin{equation} \label{eq:derivative_supp}
 - {1 \over {{{\left| {{\alpha _k}} \right|}^4}}} + {D \over {{{\left( {1 - {{\left| {{\alpha _1}} \right|}^2} - {{\left| {{\alpha _2}} \right|}^2} -  \cdots  - {{\left| {{\alpha _D}} \right|}^2}} \right)}^2}}} = 0.
\end{equation}
Solving the equations above gives the optimal probe state described by
\begin{subequations} \label{eq:probe_supp}
\begin{align}
{\left| {{\alpha _0}} \right|^2} &= {{\sqrt D } \mathord{\left/
 {\vphantom {{\sqrt D } {\left( {D + \sqrt D } \right)}}} \right.
 \kern-\nulldelimiterspace} {\left( {D + \sqrt D } \right)}}, \\
{\left| {{\alpha _k}} \right|^2} &= {1 \mathord{\left/
 {\vphantom {1 {\left( {D + \sqrt D } \right)}}} \right.
 \kern-\nulldelimiterspace} {\left( {D + \sqrt D } \right)}},{\kern 1pt} {\kern 1pt} {\kern 1pt} {\kern 1pt} {\kern 1pt} {\kern 1pt} {\kern 1pt} {\kern 1pt} {\kern 1pt} \left( \mathrm{for}~{k \ne 0} \right),
\end{align}
\end{subequations}
and a QCRB of
\begin{equation} \label{eq:SNL_supp}
{\left( {\Delta {{\bm{\theta }}_{{\rm{opt}}}}} \right)^2} = {\left( {D + \sqrt D } \right)^2}/4N.
\end{equation}

In the scenario where the parameters are defined as $\Theta_1 = {\varphi _1} = {\phi _1} - {\phi _0}, \cdots ,\Theta_D = {\varphi _D} = {\phi _{D}} - {\phi_{D-1}}$, or the phase difference between two neighboring modes, ${{\partial {f_k}\left( {\bf{\Theta }} \right)} \over {\partial {\Theta _l}}}{\rm{ = 1}}$ for $k \ge l$ and ${{\partial {f_k}\left( {\bf{\Theta }} \right)} \over {\partial {\Theta _l}}}{\rm{ = 0}}$ for $k < l$. The matrix elements of ${\cal F}^Q$ therefore become
\begin{equation} \label{eq:qfim2_supp}
{\cal F}_{l,n}^Q = 4N\left[ {\sum\limits_{k \ge \max \left( {n,l} \right)}^D {{{\left| {{\alpha _k}} \right|}^2}}  - \left( {\sum\limits_{k' \ge l}^D {{{\left| {{\alpha _{k'}}} \right|}^2}} } \right)\left( {\sum\limits_{k \ge n}^D {{{\left| {{\alpha _k}} \right|}^2}} } \right)} \right].
\end{equation}
Taking the trace of the inverse of Eq.~(\ref{eq:qfim2_supp}) gives the lower bound of ${\left( {\Delta {\bm{\varphi }}} \right)^2}$
\begin{equation}\label{eq:sql2_supp}
{\left( {\Delta {\bm{\varphi }}} \right)^2} \ge \frac{1}{N}\left({1 \over {4{{\left| {{\alpha _D}} \right|}^2}}} + {1 \over {4{{\left| {{\alpha _0}} \right|}^2}}}+ \sum\limits_{k = 1}^{D - 1} {{1 \over {2{{\left| {{\alpha _k}} \right|}^2}}}} \right).
\end{equation}
Similarly, by minimizing Eq.~(\ref{eq:sql2_supp}) under the normalization condition, the optimal probe reads
\begin{subequations} \label{eq:probe2_supp}
\begin{align}
{\left| {{\alpha _0}} \right|^2} &= {\left| {{\alpha _D}} \right|^2} = 1/\left[ {\sqrt 2 \left( {D - 1} \right) + 2} \right], \\
{\left| {{\alpha _k}} \right|^2} &= \sqrt 2 /\left[ {\sqrt 2 \left( {D - 1} \right) + 2} \right],{\kern 1pt} {\kern 1pt} {\kern 1pt} {\kern 1pt} {\kern 1pt} {\kern 1pt} {\kern 1pt} {\kern 1pt} {\kern 1pt} \left( \mathrm{for}~{k \ne 0,D} \right),
\end{align}
\end{subequations}
and the corresponding QCRB is found to be
\begin{equation} \label{eq:optimal2_supp}
{\left( {\Delta {{\bm{\varphi }}_{{\rm{opt}}}}} \right)^2} = {1 \over {4N}}{\left[ {\sqrt 2 \left( {D - 1} \right) + 2} \right]^2}.
\end{equation}

If one measures $\left\{ {{\theta _1},{\theta _2}, \cdots ,{\theta _D}} \right\}$ with the input state given by Eq.~(\ref{eq:probe_supp}) and estimates $\left\{ {{\varphi _1},{\varphi _2}, \cdots ,{\varphi _D}} \right\}$ from the measured ${\theta _k}$, the ${\left( {\Delta {\bm{\varphi }}} \right)^2}$ is bounded by~\cite{kay1993fundamentals}
\begin{eqnarray} \label{eq:delphi_supp}
{\left( {\Delta {\bm{\varphi }}} \right)^2}& \ge &{\rm{Tr}}\left[ {{\bm{J}}{{\left( {{\cal F}_{\bm{\theta }}^Q} \right)}^{ - 1}}{{\bm{J}}^T}} \right] \nonumber\\
& = &{1 \over 4N}\left[ {{{\left( {1{\rm{ + }}\sqrt D } \right)}^2} + 2\left( {D - 1} \right)\left( {\sqrt D  + D} \right)} \right],
\end{eqnarray}
where ${\bf{J}}$ is the Jacobian matrix defined as ${{\bm{J}}_{k,l}} = {{\partial {\varphi _k}\left( {\bm{\theta }} \right)} \over {\partial {\theta _l}}}$. The result of Eq.~(\ref{eq:delphi_supp}) is larger than that of Eq.~(\ref{eq:optimal2_supp}) for $D>1$. Thus it is always better to estimate $\left\{ {{\varphi _1},{\varphi _2}, \cdots ,{\varphi _D}} \right\}$ directly using the probe state given by Eq.~(\ref{eq:probe2_supp}).
\end{widetext}

\begin{thebibliography}{52}%
\makeatletter
\providecommand \@ifxundefined [1]{%
 \@ifx{#1\undefined}
}%
\providecommand \@ifnum [1]{%
 \ifnum #1\expandafter \@firstoftwo
 \else \expandafter \@secondoftwo
 \fi
}%
\providecommand \@ifx [1]{%
 \ifx #1\expandafter \@firstoftwo
 \else \expandafter \@secondoftwo
 \fi
}%
\providecommand \natexlab [1]{#1}%
\providecommand \enquote  [1]{``#1''}%
\providecommand \bibnamefont  [1]{#1}%
\providecommand \bibfnamefont [1]{#1}%
\providecommand \citenamefont [1]{#1}%
\providecommand \href@noop [0]{\@secondoftwo}%
\providecommand \href [0]{\begingroup \@sanitize@url \@href}%
\providecommand \@href[1]{\@@startlink{#1}\@@href}%
\providecommand \@@href[1]{\endgroup#1\@@endlink}%
\providecommand \@sanitize@url [0]{\catcode `\\12\catcode `\$12\catcode
  `\&12\catcode `\#12\catcode `\^12\catcode `\_12\catcode `\%12\relax}%
\providecommand \@@startlink[1]{}%
\providecommand \@@endlink[0]{}%
\providecommand \url  [0]{\begingroup\@sanitize@url \@url }%
\providecommand \@url [1]{\endgroup\@href {#1}{\urlprefix }}%
\providecommand \urlprefix  [0]{URL }%
\providecommand \Eprint [0]{\href }%
\providecommand \doibase [0]{http://dx.doi.org/}%
\providecommand \selectlanguage [0]{\@gobble}%
\providecommand \bibinfo  [0]{\@secondoftwo}%
\providecommand \bibfield  [0]{\@secondoftwo}%
\providecommand \translation [1]{[#1]}%
\providecommand \BibitemOpen [0]{}%
\providecommand \bibitemStop [0]{}%
\providecommand \bibitemNoStop [0]{.\EOS\space}%
\providecommand \EOS [0]{\spacefactor3000\relax}%
\providecommand \BibitemShut  [1]{\csname bibitem#1\endcsname}%
\let\auto@bib@innerbib\@empty
\bibitem [{\citenamefont {Giovannetti}\ \emph {et~al.}(2011)\citenamefont
  {Giovannetti}, \citenamefont {Lloyd},\ and\ \citenamefont
  {Maccone}}]{giovannetti2011advances}%
  \BibitemOpen
  \bibfield  {author} {\bibinfo {author} {\bibfnamefont {V.}~\bibnamefont
  {Giovannetti}}, \bibinfo {author} {\bibfnamefont {S.}~\bibnamefont {Lloyd}},
  \ and\ \bibinfo {author} {\bibfnamefont {L.}~\bibnamefont {Maccone}},\
  }\href@noop {} {\bibfield  {journal} {\bibinfo  {journal} {Nature Photonics}\
  }\textbf {\bibinfo {volume} {5}},\ \bibinfo {pages} {222} (\bibinfo {year}
  {2011})}\BibitemShut {NoStop}%
\bibitem [{\citenamefont {T{\'{o}}th}\ and\ \citenamefont
  {Apellaniz}(2014)}]{Toth2014information}%
  \BibitemOpen
  \bibfield  {author} {\bibinfo {author} {\bibfnamefont {G.}~\bibnamefont
  {T{\'{o}}th}}\ and\ \bibinfo {author} {\bibfnamefont {I.}~\bibnamefont
  {Apellaniz}},\ }\href@noop {} {\bibfield  {journal} {\bibinfo  {journal}
  {Journal of Physics A: Mathematical and Theoretical}\ }\textbf {\bibinfo
  {volume} {47}},\ \bibinfo {pages} {424006} (\bibinfo {year}
  {2014})}\BibitemShut {NoStop}%
\bibitem [{\citenamefont {Degen}\ \emph {et~al.}(2017)\citenamefont {Degen},
  \citenamefont {Reinhard},\ and\ \citenamefont
  {Cappellaro}}]{degen2017quantum}%
  \BibitemOpen
  \bibfield  {author} {\bibinfo {author} {\bibfnamefont {C.~L.}\ \bibnamefont
  {Degen}}, \bibinfo {author} {\bibfnamefont {F.}~\bibnamefont {Reinhard}}, \
  and\ \bibinfo {author} {\bibfnamefont {P.}~\bibnamefont {Cappellaro}},\
  }\href@noop {} {\bibfield  {journal} {\bibinfo  {journal} {Rev. Mod. Phys.}\
  }\textbf {\bibinfo {volume} {89}},\ \bibinfo {pages} {035002} (\bibinfo
  {year} {2017})}\BibitemShut {NoStop}%
\bibitem [{\citenamefont {Pezz{\`e}}\ \emph {et~al.}(2018)\citenamefont
  {Pezz{\`e}}, \citenamefont {Smerzi}, \citenamefont {Oberthaler},
  \citenamefont {Schmied},\ and\ \citenamefont {Treutlein}}]{pezze2018quantum}%
  \BibitemOpen
  \bibfield  {author} {\bibinfo {author} {\bibfnamefont {L.}~\bibnamefont
  {Pezz{\`e}}}, \bibinfo {author} {\bibfnamefont {A.}~\bibnamefont {Smerzi}},
  \bibinfo {author} {\bibfnamefont {M.~K.}\ \bibnamefont {Oberthaler}},
  \bibinfo {author} {\bibfnamefont {R.}~\bibnamefont {Schmied}}, \ and\
  \bibinfo {author} {\bibfnamefont {P.}~\bibnamefont {Treutlein}},\ }\href@noop
  {} {\bibfield  {journal} {\bibinfo  {journal} {Reviews of Modern Physics}\
  }\textbf {\bibinfo {volume} {90}},\ \bibinfo {pages} {035005} (\bibinfo
  {year} {2018})}\BibitemShut {NoStop}%
\bibitem [{\citenamefont {Giovannetti}\ \emph {et~al.}(2006)\citenamefont
  {Giovannetti}, \citenamefont {Lloyd},\ and\ \citenamefont
  {Maccone}}]{giovannetti2006quantum}%
  \BibitemOpen
  \bibfield  {author} {\bibinfo {author} {\bibfnamefont {V.}~\bibnamefont
  {Giovannetti}}, \bibinfo {author} {\bibfnamefont {S.}~\bibnamefont {Lloyd}},
  \ and\ \bibinfo {author} {\bibfnamefont {L.}~\bibnamefont {Maccone}},\
  }\href@noop {} {\bibfield  {journal} {\bibinfo  {journal} {Physical review
  letters}\ }\textbf {\bibinfo {volume} {96}},\ \bibinfo {pages} {010401}
  (\bibinfo {year} {2006})}\BibitemShut {NoStop}%
\bibitem [{\citenamefont {Szczykulska}\ \emph {et~al.}(2016)\citenamefont
  {Szczykulska}, \citenamefont {Baumgratz},\ and\ \citenamefont
  {Datta}}]{szczykulska2016multi}%
  \BibitemOpen
  \bibfield  {author} {\bibinfo {author} {\bibfnamefont {M.}~\bibnamefont
  {Szczykulska}}, \bibinfo {author} {\bibfnamefont {T.}~\bibnamefont
  {Baumgratz}}, \ and\ \bibinfo {author} {\bibfnamefont {A.}~\bibnamefont
  {Datta}},\ }\href@noop {} {\bibfield  {journal} {\bibinfo  {journal}
  {Advances in Physics: X}\ }\textbf {\bibinfo {volume} {1}},\ \bibinfo {pages}
  {621} (\bibinfo {year} {2016})}\BibitemShut {NoStop}%
\bibitem [{\citenamefont {Spagnolo}\ \emph {et~al.}(2012)\citenamefont
  {Spagnolo}, \citenamefont {Aparo}, \citenamefont {Vitelli}, \citenamefont
  {Crespi}, \citenamefont {Ramponi}, \citenamefont {Osellame}, \citenamefont
  {Mataloni},\ and\ \citenamefont {Sciarrino}}]{spagnolo2012quantum}%
  \BibitemOpen
  \bibfield  {author} {\bibinfo {author} {\bibfnamefont {N.}~\bibnamefont
  {Spagnolo}}, \bibinfo {author} {\bibfnamefont {L.}~\bibnamefont {Aparo}},
  \bibinfo {author} {\bibfnamefont {C.}~\bibnamefont {Vitelli}}, \bibinfo
  {author} {\bibfnamefont {A.}~\bibnamefont {Crespi}}, \bibinfo {author}
  {\bibfnamefont {R.}~\bibnamefont {Ramponi}}, \bibinfo {author} {\bibfnamefont
  {R.}~\bibnamefont {Osellame}}, \bibinfo {author} {\bibfnamefont
  {P.}~\bibnamefont {Mataloni}}, \ and\ \bibinfo {author} {\bibfnamefont
  {F.}~\bibnamefont {Sciarrino}},\ }\href@noop {} {\bibfield  {journal}
  {\bibinfo  {journal} {Scientific Reports}\ }\textbf {\bibinfo {volume} {2}},\
  \bibinfo {pages} {862} (\bibinfo {year} {2012})}\BibitemShut {NoStop}%
\bibitem [{\citenamefont {Humphreys}\ \emph {et~al.}(2013)\citenamefont
  {Humphreys}, \citenamefont {Barbieri}, \citenamefont {Datta},\ and\
  \citenamefont {Walmsley}}]{humphreys2013quantum}%
  \BibitemOpen
  \bibfield  {author} {\bibinfo {author} {\bibfnamefont {P.~C.}\ \bibnamefont
  {Humphreys}}, \bibinfo {author} {\bibfnamefont {M.}~\bibnamefont {Barbieri}},
  \bibinfo {author} {\bibfnamefont {A.}~\bibnamefont {Datta}}, \ and\ \bibinfo
  {author} {\bibfnamefont {I.~A.}\ \bibnamefont {Walmsley}},\ }\href@noop {}
  {\bibfield  {journal} {\bibinfo  {journal} {Physical Review Letters}\
  }\textbf {\bibinfo {volume} {111}},\ \bibinfo {pages} {070403} (\bibinfo
  {year} {2013})}\BibitemShut {NoStop}%
\bibitem [{\citenamefont {Pinel}\ \emph {et~al.}(2013)\citenamefont {Pinel},
  \citenamefont {Jian}, \citenamefont {Treps}, \citenamefont {Fabre},\ and\
  \citenamefont {Braun}}]{pinel2013quantum}%
  \BibitemOpen
  \bibfield  {author} {\bibinfo {author} {\bibfnamefont {O.}~\bibnamefont
  {Pinel}}, \bibinfo {author} {\bibfnamefont {P.}~\bibnamefont {Jian}},
  \bibinfo {author} {\bibfnamefont {N.}~\bibnamefont {Treps}}, \bibinfo
  {author} {\bibfnamefont {C.}~\bibnamefont {Fabre}}, \ and\ \bibinfo {author}
  {\bibfnamefont {D.}~\bibnamefont {Braun}},\ }\href {\doibase
  10.1103/PhysRevA.88.040102} {\bibfield  {journal} {\bibinfo  {journal} {Phys.
  Rev. A}\ }\textbf {\bibinfo {volume} {88}},\ \bibinfo {pages} {040102}
  (\bibinfo {year} {2013})}\BibitemShut {NoStop}%
\bibitem [{\citenamefont {Genoni}\ \emph {et~al.}(2013)\citenamefont {Genoni},
  \citenamefont {Paris}, \citenamefont {Adesso}, \citenamefont {Nha},
  \citenamefont {Knight},\ and\ \citenamefont {Kim}}]{kim2013joint}%
  \BibitemOpen
  \bibfield  {author} {\bibinfo {author} {\bibfnamefont {M.~G.}\ \bibnamefont
  {Genoni}}, \bibinfo {author} {\bibfnamefont {M.~G.~A.}\ \bibnamefont
  {Paris}}, \bibinfo {author} {\bibfnamefont {G.}~\bibnamefont {Adesso}},
  \bibinfo {author} {\bibfnamefont {H.}~\bibnamefont {Nha}}, \bibinfo {author}
  {\bibfnamefont {P.~L.}\ \bibnamefont {Knight}}, \ and\ \bibinfo {author}
  {\bibfnamefont {M.~S.}\ \bibnamefont {Kim}},\ }\href@noop {} {\bibfield
  {journal} {\bibinfo  {journal} {Physical Review A}\ }\textbf {\bibinfo
  {volume} {87}},\ \bibinfo {pages} {012107} (\bibinfo {year}
  {2013})}\BibitemShut {NoStop}%
\bibitem [{\citenamefont {Vidrighin}\ \emph {et~al.}(2014)\citenamefont
  {Vidrighin}, \citenamefont {Donati}, \citenamefont {Genoni}, \citenamefont
  {Jin}, \citenamefont {Kolthammer}, \citenamefont {Kim}, \citenamefont
  {Datta}, \citenamefont {Barbieri},\ and\ \citenamefont
  {Walmsley}}]{vidrighin2014joint}%
  \BibitemOpen
  \bibfield  {author} {\bibinfo {author} {\bibfnamefont {M.~D.}\ \bibnamefont
  {Vidrighin}}, \bibinfo {author} {\bibfnamefont {G.}~\bibnamefont {Donati}},
  \bibinfo {author} {\bibfnamefont {M.~G.}\ \bibnamefont {Genoni}}, \bibinfo
  {author} {\bibfnamefont {X.-M.}\ \bibnamefont {Jin}}, \bibinfo {author}
  {\bibfnamefont {W.~S.}\ \bibnamefont {Kolthammer}}, \bibinfo {author}
  {\bibfnamefont {M.}~\bibnamefont {Kim}}, \bibinfo {author} {\bibfnamefont
  {A.}~\bibnamefont {Datta}}, \bibinfo {author} {\bibfnamefont
  {M.}~\bibnamefont {Barbieri}}, \ and\ \bibinfo {author} {\bibfnamefont
  {I.~A.}\ \bibnamefont {Walmsley}},\ }\href@noop {} {\bibfield  {journal}
  {\bibinfo  {journal} {Nature Communications}\ }\textbf {\bibinfo {volume}
  {5}},\ \bibinfo {pages} {3532} (\bibinfo {year} {2014})}\BibitemShut
  {NoStop}%
\bibitem [{\citenamefont {Crowley}\ \emph {et~al.}(2014)\citenamefont
  {Crowley}, \citenamefont {Datta}, \citenamefont {Barbieri},\ and\
  \citenamefont {Walmsley}}]{crowley2014tradeoff}%
  \BibitemOpen
  \bibfield  {author} {\bibinfo {author} {\bibfnamefont {P.~J.~D.}\
  \bibnamefont {Crowley}}, \bibinfo {author} {\bibfnamefont {A.}~\bibnamefont
  {Datta}}, \bibinfo {author} {\bibfnamefont {M.}~\bibnamefont {Barbieri}}, \
  and\ \bibinfo {author} {\bibfnamefont {I.~A.}\ \bibnamefont {Walmsley}},\
  }\href {\doibase 10.1103/PhysRevA.89.023845} {\bibfield  {journal} {\bibinfo
  {journal} {Phys. Rev. A}\ }\textbf {\bibinfo {volume} {89}},\ \bibinfo
  {pages} {023845} (\bibinfo {year} {2014})}\BibitemShut {NoStop}%
\bibitem [{\citenamefont {Yao}\ \emph {et~al.}(2014)\citenamefont {Yao},
  \citenamefont {Ge}, \citenamefont {Xiao}, \citenamefont {Wang},\ and\
  \citenamefont {Sun}}]{yao2014multiple}%
  \BibitemOpen
  \bibfield  {author} {\bibinfo {author} {\bibfnamefont {Y.}~\bibnamefont
  {Yao}}, \bibinfo {author} {\bibfnamefont {L.}~\bibnamefont {Ge}}, \bibinfo
  {author} {\bibfnamefont {X.}~\bibnamefont {Xiao}}, \bibinfo {author}
  {\bibfnamefont {X.-g.}\ \bibnamefont {Wang}}, \ and\ \bibinfo {author}
  {\bibfnamefont {C.-p.}\ \bibnamefont {Sun}},\ }\href@noop {} {\bibfield
  {journal} {\bibinfo  {journal} {Physical Review A}\ }\textbf {\bibinfo
  {volume} {90}},\ \bibinfo {pages} {022327} (\bibinfo {year}
  {2014})}\BibitemShut {NoStop}%
\bibitem [{\citenamefont {Zhang}\ and\ \citenamefont
  {Fan}(2014)}]{zhang2014quantum}%
  \BibitemOpen
  \bibfield  {author} {\bibinfo {author} {\bibfnamefont {Y.-R.}\ \bibnamefont
  {Zhang}}\ and\ \bibinfo {author} {\bibfnamefont {H.}~\bibnamefont {Fan}},\
  }\href@noop {} {\bibfield  {journal} {\bibinfo  {journal} {Physical Review
  A}\ }\textbf {\bibinfo {volume} {90}},\ \bibinfo {pages} {043818} (\bibinfo
  {year} {2014})}\BibitemShut {NoStop}%
\bibitem [{\citenamefont {Yue}\ \emph {et~al.}(2014)\citenamefont {Yue},
  \citenamefont {Zhang},\ and\ \citenamefont {Fan}}]{yue2014quantum}%
  \BibitemOpen
  \bibfield  {author} {\bibinfo {author} {\bibfnamefont {J.-D.}\ \bibnamefont
  {Yue}}, \bibinfo {author} {\bibfnamefont {Y.-R.}\ \bibnamefont {Zhang}}, \
  and\ \bibinfo {author} {\bibfnamefont {H.}~\bibnamefont {Fan}},\ }\href@noop
  {} {\bibfield  {journal} {\bibinfo  {journal} {Scientific Reports}\ }\textbf
  {\bibinfo {volume} {4}},\ \bibinfo {pages} {5933} (\bibinfo {year}
  {2014})}\BibitemShut {NoStop}%
\bibitem [{\citenamefont {Berry}\ \emph {et~al.}(2015)\citenamefont {Berry},
  \citenamefont {Tsang}, \citenamefont {Hall},\ and\ \citenamefont
  {Wiseman}}]{berry2015quantum}%
  \BibitemOpen
  \bibfield  {author} {\bibinfo {author} {\bibfnamefont {D.~W.}\ \bibnamefont
  {Berry}}, \bibinfo {author} {\bibfnamefont {M.}~\bibnamefont {Tsang}},
  \bibinfo {author} {\bibfnamefont {M.~J.~W.}\ \bibnamefont {Hall}}, \ and\
  \bibinfo {author} {\bibfnamefont {H.~M.}\ \bibnamefont {Wiseman}},\ }\href
  {\doibase 10.1103/PhysRevX.5.031018} {\bibfield  {journal} {\bibinfo
  {journal} {Phys. Rev. X}\ }\textbf {\bibinfo {volume} {5}},\ \bibinfo {pages}
  {031018} (\bibinfo {year} {2015})}\BibitemShut {NoStop}%
\bibitem [{\citenamefont {Liu}\ \emph {et~al.}(2015)\citenamefont {Liu},
  \citenamefont {Jing},\ and\ \citenamefont {Wang}}]{liu2015quantum}%
  \BibitemOpen
  \bibfield  {author} {\bibinfo {author} {\bibfnamefont {J.}~\bibnamefont
  {Liu}}, \bibinfo {author} {\bibfnamefont {X.-X.}\ \bibnamefont {Jing}}, \
  and\ \bibinfo {author} {\bibfnamefont {X.}~\bibnamefont {Wang}},\ }\href@noop
  {} {\bibfield  {journal} {\bibinfo  {journal} {Scientific Reports}\ }\textbf
  {\bibinfo {volume} {5}},\ \bibinfo {pages} {8565} (\bibinfo {year}
  {2015})}\BibitemShut {NoStop}%
\bibitem [{\citenamefont {Baumgratz}\ and\ \citenamefont
  {Datta}(2016)}]{baumgratz2016quantum}%
  \BibitemOpen
  \bibfield  {author} {\bibinfo {author} {\bibfnamefont {T.}~\bibnamefont
  {Baumgratz}}\ and\ \bibinfo {author} {\bibfnamefont {A.}~\bibnamefont
  {Datta}},\ }\href@noop {} {\bibfield  {journal} {\bibinfo  {journal}
  {Physical Review Letters}\ }\textbf {\bibinfo {volume} {116}},\ \bibinfo
  {pages} {030801} (\bibinfo {year} {2016})}\BibitemShut {NoStop}%
\bibitem [{\citenamefont {Gagatsos}\ \emph {et~al.}(2016)\citenamefont
  {Gagatsos}, \citenamefont {Branford},\ and\ \citenamefont
  {Datta}}]{gagatsos2016gaussian}%
  \BibitemOpen
  \bibfield  {author} {\bibinfo {author} {\bibfnamefont {C.~N.}\ \bibnamefont
  {Gagatsos}}, \bibinfo {author} {\bibfnamefont {D.}~\bibnamefont {Branford}},
  \ and\ \bibinfo {author} {\bibfnamefont {A.}~\bibnamefont {Datta}},\
  }\href@noop {} {\bibfield  {journal} {\bibinfo  {journal} {Physical Review
  A}\ }\textbf {\bibinfo {volume} {94}},\ \bibinfo {pages} {042342} (\bibinfo
  {year} {2016})}\BibitemShut {NoStop}%
\bibitem [{\citenamefont {Ragy}\ \emph {et~al.}(2016)\citenamefont {Ragy},
  \citenamefont {Jarzyna},\ and\ \citenamefont
  {Demkowicz-Dobrza{\'n}ski}}]{ragy2016compatibility}%
  \BibitemOpen
  \bibfield  {author} {\bibinfo {author} {\bibfnamefont {S.}~\bibnamefont
  {Ragy}}, \bibinfo {author} {\bibfnamefont {M.}~\bibnamefont {Jarzyna}}, \
  and\ \bibinfo {author} {\bibfnamefont {R.}~\bibnamefont
  {Demkowicz-Dobrza{\'n}ski}},\ }\href@noop {} {\bibfield  {journal} {\bibinfo
  {journal} {Physical Review A}\ }\textbf {\bibinfo {volume} {94}},\ \bibinfo
  {pages} {052108} (\bibinfo {year} {2016})}\BibitemShut {NoStop}%
\bibitem [{\citenamefont {Knott}\ \emph {et~al.}(2016)\citenamefont {Knott},
  \citenamefont {Proctor}, \citenamefont {Hayes}, \citenamefont {Ralph},
  \citenamefont {Kok},\ and\ \citenamefont {Dunningham}}]{knott2016local}%
  \BibitemOpen
  \bibfield  {author} {\bibinfo {author} {\bibfnamefont {P.~A.}\ \bibnamefont
  {Knott}}, \bibinfo {author} {\bibfnamefont {T.~J.}\ \bibnamefont {Proctor}},
  \bibinfo {author} {\bibfnamefont {A.~J.}\ \bibnamefont {Hayes}}, \bibinfo
  {author} {\bibfnamefont {J.~F.}\ \bibnamefont {Ralph}}, \bibinfo {author}
  {\bibfnamefont {P.}~\bibnamefont {Kok}}, \ and\ \bibinfo {author}
  {\bibfnamefont {J.~A.}\ \bibnamefont {Dunningham}},\ }\href {\doibase
  10.1103/PhysRevA.94.062312} {\bibfield  {journal} {\bibinfo  {journal} {Phys.
  Rev. A}\ }\textbf {\bibinfo {volume} {94}},\ \bibinfo {pages} {062312}
  (\bibinfo {year} {2016})}\BibitemShut {NoStop}%
\bibitem [{\citenamefont {Ciampini}\ \emph {et~al.}(2016)\citenamefont
  {Ciampini}, \citenamefont {Spagnolo}, \citenamefont {Vitelli}, \citenamefont
  {Pezz{\`e}}, \citenamefont {Smerzi},\ and\ \citenamefont
  {Sciarrino}}]{ciampini2016quantum}%
  \BibitemOpen
  \bibfield  {author} {\bibinfo {author} {\bibfnamefont {M.~A.}\ \bibnamefont
  {Ciampini}}, \bibinfo {author} {\bibfnamefont {N.}~\bibnamefont {Spagnolo}},
  \bibinfo {author} {\bibfnamefont {C.}~\bibnamefont {Vitelli}}, \bibinfo
  {author} {\bibfnamefont {L.}~\bibnamefont {Pezz{\`e}}}, \bibinfo {author}
  {\bibfnamefont {A.}~\bibnamefont {Smerzi}}, \ and\ \bibinfo {author}
  {\bibfnamefont {F.}~\bibnamefont {Sciarrino}},\ }\href@noop {} {\bibfield
  {journal} {\bibinfo  {journal} {Scientific Reports}\ }\textbf {\bibinfo
  {volume} {6}},\ \bibinfo {pages} {28881} (\bibinfo {year}
  {2016})}\BibitemShut {NoStop}%
\bibitem [{\citenamefont {Liu}\ \emph {et~al.}(2016)\citenamefont {Liu},
  \citenamefont {Lu}, \citenamefont {Sun},\ and\ \citenamefont
  {Wang}}]{liu2016quantum}%
  \BibitemOpen
  \bibfield  {author} {\bibinfo {author} {\bibfnamefont {J.}~\bibnamefont
  {Liu}}, \bibinfo {author} {\bibfnamefont {X.-M.}\ \bibnamefont {Lu}},
  \bibinfo {author} {\bibfnamefont {Z.}~\bibnamefont {Sun}}, \ and\ \bibinfo
  {author} {\bibfnamefont {X.}~\bibnamefont {Wang}},\ }\href@noop {} {\bibfield
   {journal} {\bibinfo  {journal} {Journal of Physics A: Mathematical and
  Theoretical}\ }\textbf {\bibinfo {volume} {49}},\ \bibinfo {pages} {115302}
  (\bibinfo {year} {2016})}\BibitemShut {NoStop}%
\bibitem [{\citenamefont {Kok}\ \emph {et~al.}(2017)\citenamefont {Kok},
  \citenamefont {Dunningham},\ and\ \citenamefont {Ralph}}]{kok2017role}%
  \BibitemOpen
  \bibfield  {author} {\bibinfo {author} {\bibfnamefont {P.}~\bibnamefont
  {Kok}}, \bibinfo {author} {\bibfnamefont {J.}~\bibnamefont {Dunningham}}, \
  and\ \bibinfo {author} {\bibfnamefont {J.~F.}\ \bibnamefont {Ralph}},\
  }\href@noop {} {\bibfield  {journal} {\bibinfo  {journal} {Physical Review
  A}\ }\textbf {\bibinfo {volume} {95}},\ \bibinfo {pages} {012326} (\bibinfo
  {year} {2017})}\BibitemShut {NoStop}%
\bibitem [{\citenamefont {Yousefjani}\ \emph {et~al.}(2017)\citenamefont
  {Yousefjani}, \citenamefont {Nichols}, \citenamefont {Salimi},\ and\
  \citenamefont {Adesso}}]{yousefjani2017estimating}%
  \BibitemOpen
  \bibfield  {author} {\bibinfo {author} {\bibfnamefont {R.}~\bibnamefont
  {Yousefjani}}, \bibinfo {author} {\bibfnamefont {R.}~\bibnamefont {Nichols}},
  \bibinfo {author} {\bibfnamefont {S.}~\bibnamefont {Salimi}}, \ and\ \bibinfo
  {author} {\bibfnamefont {G.}~\bibnamefont {Adesso}},\ }\href@noop {}
  {\bibfield  {journal} {\bibinfo  {journal} {Physical Review A}\ }\textbf
  {\bibinfo {volume} {95}},\ \bibinfo {pages} {062307} (\bibinfo {year}
  {2017})}\BibitemShut {NoStop}%
\bibitem [{\citenamefont {Liu}\ and\ \citenamefont
  {Yuan}(2017)}]{liu2017control}%
  \BibitemOpen
  \bibfield  {author} {\bibinfo {author} {\bibfnamefont {J.}~\bibnamefont
  {Liu}}\ and\ \bibinfo {author} {\bibfnamefont {H.}~\bibnamefont {Yuan}},\
  }\href@noop {} {\bibfield  {journal} {\bibinfo  {journal} {Physical Review
  A}\ }\textbf {\bibinfo {volume} {96}},\ \bibinfo {pages} {042114} (\bibinfo
  {year} {2017})}\BibitemShut {NoStop}%
\bibitem [{\citenamefont {Pezz{\`e}}\ \emph {et~al.}(2017)\citenamefont
  {Pezz{\`e}}, \citenamefont {Ciampini}, \citenamefont {Spagnolo},
  \citenamefont {Humphreys}, \citenamefont {Datta}, \citenamefont {Walmsley},
  \citenamefont {Barbieri}, \citenamefont {Sciarrino},\ and\ \citenamefont
  {Smerzi}}]{pezze2017optimal}%
  \BibitemOpen
  \bibfield  {author} {\bibinfo {author} {\bibfnamefont {L.}~\bibnamefont
  {Pezz{\`e}}}, \bibinfo {author} {\bibfnamefont {M.~A.}\ \bibnamefont
  {Ciampini}}, \bibinfo {author} {\bibfnamefont {N.}~\bibnamefont {Spagnolo}},
  \bibinfo {author} {\bibfnamefont {P.~C.}\ \bibnamefont {Humphreys}}, \bibinfo
  {author} {\bibfnamefont {A.}~\bibnamefont {Datta}}, \bibinfo {author}
  {\bibfnamefont {I.~A.}\ \bibnamefont {Walmsley}}, \bibinfo {author}
  {\bibfnamefont {M.}~\bibnamefont {Barbieri}}, \bibinfo {author}
  {\bibfnamefont {F.}~\bibnamefont {Sciarrino}}, \ and\ \bibinfo {author}
  {\bibfnamefont {A.}~\bibnamefont {Smerzi}},\ }\href@noop {} {\bibfield
  {journal} {\bibinfo  {journal} {Physical Review Letters}\ }\textbf {\bibinfo
  {volume} {119}},\ \bibinfo {pages} {130504} (\bibinfo {year}
  {2017})}\BibitemShut {NoStop}%
\bibitem [{\citenamefont {\ifmmode \check{R}\else
  \v{R}\fi{}eha\ifmmode~\check{c}\else \v{c}\fi{}ek}\ \emph
  {et~al.}(2017)\citenamefont {\ifmmode \check{R}\else
  \v{R}\fi{}eha\ifmmode~\check{c}\else \v{c}\fi{}ek}, \citenamefont {Hradil},
  \citenamefont {Stoklasa}, \citenamefont {Pa\'ur}, \citenamefont {Grover},
  \citenamefont {Krzic},\ and\ \citenamefont
  {S\'anchez-Soto}}]{vrehavcek2017multiparameter}%
  \BibitemOpen
  \bibfield  {author} {\bibinfo {author} {\bibfnamefont {J.}~\bibnamefont
  {\ifmmode \check{R}\else \v{R}\fi{}eha\ifmmode~\check{c}\else \v{c}\fi{}ek}},
  \bibinfo {author} {\bibfnamefont {Z.}~\bibnamefont {Hradil}}, \bibinfo
  {author} {\bibfnamefont {B.}~\bibnamefont {Stoklasa}}, \bibinfo {author}
  {\bibfnamefont {M.}~\bibnamefont {Pa\'ur}}, \bibinfo {author} {\bibfnamefont
  {J.}~\bibnamefont {Grover}}, \bibinfo {author} {\bibfnamefont
  {A.}~\bibnamefont {Krzic}}, \ and\ \bibinfo {author} {\bibfnamefont {L.~L.}\
  \bibnamefont {S\'anchez-Soto}},\ }\href {\doibase 10.1103/PhysRevA.96.062107}
  {\bibfield  {journal} {\bibinfo  {journal} {Phys. Rev. A}\ }\textbf {\bibinfo
  {volume} {96}},\ \bibinfo {pages} {062107} (\bibinfo {year}
  {2017})}\BibitemShut {NoStop}%
\bibitem [{\citenamefont {Zhang}\ and\ \citenamefont
  {Chan}(2017)}]{zhang2017quantum}%
  \BibitemOpen
  \bibfield  {author} {\bibinfo {author} {\bibfnamefont {L.}~\bibnamefont
  {Zhang}}\ and\ \bibinfo {author} {\bibfnamefont {K.~W.~C.}\ \bibnamefont
  {Chan}},\ }\href {\doibase 10.1103/PhysRevA.95.032321} {\bibfield  {journal}
  {\bibinfo  {journal} {Phys. Rev. A}\ }\textbf {\bibinfo {volume} {95}},\
  \bibinfo {pages} {032321} (\bibinfo {year} {2017})}\BibitemShut {NoStop}%
\bibitem [{\citenamefont {Zhuang}\ \emph {et~al.}(2017)\citenamefont {Zhuang},
  \citenamefont {Zhang},\ and\ \citenamefont
  {Shapiro}}]{zhuang2017entanglement}%
  \BibitemOpen
  \bibfield  {author} {\bibinfo {author} {\bibfnamefont {Q.}~\bibnamefont
  {Zhuang}}, \bibinfo {author} {\bibfnamefont {Z.}~\bibnamefont {Zhang}}, \
  and\ \bibinfo {author} {\bibfnamefont {J.~H.}\ \bibnamefont {Shapiro}},\
  }\href {\doibase 10.1103/PhysRevA.96.040304} {\bibfield  {journal} {\bibinfo
  {journal} {Phys. Rev. A}\ }\textbf {\bibinfo {volume} {96}},\ \bibinfo
  {pages} {040304} (\bibinfo {year} {2017})}\BibitemShut {NoStop}%
\bibitem [{\citenamefont {Kura}\ and\ \citenamefont
  {Ueda}(2018)}]{kura2018finite}%
  \BibitemOpen
  \bibfield  {author} {\bibinfo {author} {\bibfnamefont {N.}~\bibnamefont
  {Kura}}\ and\ \bibinfo {author} {\bibfnamefont {M.}~\bibnamefont {Ueda}},\
  }\href@noop {} {\bibfield  {journal} {\bibinfo  {journal} {Physical Review
  A}\ }\textbf {\bibinfo {volume} {97}},\ \bibinfo {pages} {012101} (\bibinfo
  {year} {2018})}\BibitemShut {NoStop}%
\bibitem [{\citenamefont {Li}\ \emph {et~al.}(2018{\natexlab{a}})\citenamefont
  {Li}, \citenamefont {Liu}, \citenamefont {Cui}, \citenamefont {Huo},
  \citenamefont {Assad}, \citenamefont {Li},\ and\ \citenamefont
  {Ou}}]{zyou2018joint}%
  \BibitemOpen
  \bibfield  {author} {\bibinfo {author} {\bibfnamefont {J.}~\bibnamefont
  {Li}}, \bibinfo {author} {\bibfnamefont {Y.}~\bibnamefont {Liu}}, \bibinfo
  {author} {\bibfnamefont {L.}~\bibnamefont {Cui}}, \bibinfo {author}
  {\bibfnamefont {N.}~\bibnamefont {Huo}}, \bibinfo {author} {\bibfnamefont
  {S.~M.}\ \bibnamefont {Assad}}, \bibinfo {author} {\bibfnamefont
  {X.}~\bibnamefont {Li}}, \ and\ \bibinfo {author} {\bibfnamefont {Z.~Y.}\
  \bibnamefont {Ou}},\ }\href@noop {} {\bibfield  {journal} {\bibinfo
  {journal} {Physical Review A}\ }\textbf {\bibinfo {volume} {97}},\ \bibinfo
  {pages} {052127} (\bibinfo {year} {2018}{\natexlab{a}})}\BibitemShut
  {NoStop}%
\bibitem [{\citenamefont {Li}\ \emph {et~al.}(2018{\natexlab{b}})\citenamefont
  {Li}, \citenamefont {Liu}, \citenamefont {Cui}, \citenamefont {Huo},
  \citenamefont {Assad}, \citenamefont {Li},\ and\ \citenamefont
  {Ou}}]{liu2018loss}%
  \BibitemOpen
  \bibfield  {author} {\bibinfo {author} {\bibfnamefont {J.}~\bibnamefont
  {Li}}, \bibinfo {author} {\bibfnamefont {Y.}~\bibnamefont {Liu}}, \bibinfo
  {author} {\bibfnamefont {L.}~\bibnamefont {Cui}}, \bibinfo {author}
  {\bibfnamefont {N.}~\bibnamefont {Huo}}, \bibinfo {author} {\bibfnamefont
  {S.~M.}\ \bibnamefont {Assad}}, \bibinfo {author} {\bibfnamefont
  {X.}~\bibnamefont {Li}}, \ and\ \bibinfo {author} {\bibfnamefont {Z.~Y.}\
  \bibnamefont {Ou}},\ }\href@noop {} {\bibfield  {journal} {\bibinfo
  {journal} {Optics Express}\ }\textbf {\bibinfo {volume} {26}},\ \bibinfo
  {pages} {27705} (\bibinfo {year} {2018}{\natexlab{b}})}\BibitemShut {NoStop}%
\bibitem [{\citenamefont {Bradshaw}\ \emph {et~al.}(2018)\citenamefont
  {Bradshaw}, \citenamefont {Lam},\ and\ \citenamefont
  {Assad}}]{bradshaw2018ultimate}%
  \BibitemOpen
  \bibfield  {author} {\bibinfo {author} {\bibfnamefont {M.}~\bibnamefont
  {Bradshaw}}, \bibinfo {author} {\bibfnamefont {P.~K.}\ \bibnamefont {Lam}}, \
  and\ \bibinfo {author} {\bibfnamefont {S.~M.}\ \bibnamefont {Assad}},\
  }\href@noop {} {\bibfield  {journal} {\bibinfo  {journal} {Physical Review
  A}\ }\textbf {\bibinfo {volume} {97}},\ \bibinfo {pages} {012106} (\bibinfo
  {year} {2018})}\BibitemShut {NoStop}%
\bibitem [{\citenamefont {Nichols}\ \emph {et~al.}(2018)\citenamefont
  {Nichols}, \citenamefont {Liuzzo-Scorpo}, \citenamefont {Knott},\ and\
  \citenamefont {Adesso}}]{nichols2018multiparameter}%
  \BibitemOpen
  \bibfield  {author} {\bibinfo {author} {\bibfnamefont {R.}~\bibnamefont
  {Nichols}}, \bibinfo {author} {\bibfnamefont {P.}~\bibnamefont
  {Liuzzo-Scorpo}}, \bibinfo {author} {\bibfnamefont {P.~A.}\ \bibnamefont
  {Knott}}, \ and\ \bibinfo {author} {\bibfnamefont {G.}~\bibnamefont
  {Adesso}},\ }\href@noop {} {\bibfield  {journal} {\bibinfo  {journal}
  {Physical Review A}\ }\textbf {\bibinfo {volume} {98}},\ \bibinfo {pages}
  {012114} (\bibinfo {year} {2018})}\BibitemShut {NoStop}%
\bibitem [{\citenamefont {Zhuang}\ \emph {et~al.}(2018)\citenamefont {Zhuang},
  \citenamefont {Huang},\ and\ \citenamefont {Lee}}]{zhuang2018multiparameter}%
  \BibitemOpen
  \bibfield  {author} {\bibinfo {author} {\bibfnamefont {M.}~\bibnamefont
  {Zhuang}}, \bibinfo {author} {\bibfnamefont {J.}~\bibnamefont {Huang}}, \
  and\ \bibinfo {author} {\bibfnamefont {C.}~\bibnamefont {Lee}},\ }\href@noop
  {} {\bibfield  {journal} {\bibinfo  {journal} {Physical Review A}\ }\textbf
  {\bibinfo {volume} {98}},\ \bibinfo {pages} {033603} (\bibinfo {year}
  {2018})}\BibitemShut {NoStop}%
\bibitem [{\citenamefont {Proctor}\ \emph {et~al.}(2018)\citenamefont
  {Proctor}, \citenamefont {Knott},\ and\ \citenamefont
  {Dunningham}}]{proctor2018multiparameter}%
  \BibitemOpen
  \bibfield  {author} {\bibinfo {author} {\bibfnamefont {T.~J.}\ \bibnamefont
  {Proctor}}, \bibinfo {author} {\bibfnamefont {P.~A.}\ \bibnamefont {Knott}},
  \ and\ \bibinfo {author} {\bibfnamefont {J.~A.}\ \bibnamefont {Dunningham}},\
  }\href@noop {} {\bibfield  {journal} {\bibinfo  {journal} {Physical Review
  Letters}\ }\textbf {\bibinfo {volume} {120}},\ \bibinfo {pages} {080501}
  (\bibinfo {year} {2018})}\BibitemShut {NoStop}%
\bibitem [{\citenamefont {Gessner}\ \emph {et~al.}(2018)\citenamefont
  {Gessner}, \citenamefont {Pezz{\`e}},\ and\ \citenamefont
  {Smerzi}}]{gessner2018sensitivity}%
  \BibitemOpen
  \bibfield  {author} {\bibinfo {author} {\bibfnamefont {M.}~\bibnamefont
  {Gessner}}, \bibinfo {author} {\bibfnamefont {L.}~\bibnamefont {Pezz{\`e}}},
  \ and\ \bibinfo {author} {\bibfnamefont {A.}~\bibnamefont {Smerzi}},\
  }\href@noop {} {\bibfield  {journal} {\bibinfo  {journal} {Physical Review
  Letters}\ }\textbf {\bibinfo {volume} {121}},\ \bibinfo {pages} {130503}
  (\bibinfo {year} {2018})}\BibitemShut {NoStop}%
\bibitem [{\citenamefont {Yang}\ \emph {et~al.}(2019)\citenamefont {Yang},
  \citenamefont {Pang}, \citenamefont {Zhou},\ and\ \citenamefont
  {Jordan}}]{yang2018optimal}%
  \BibitemOpen
  \bibfield  {author} {\bibinfo {author} {\bibfnamefont {J.}~\bibnamefont
  {Yang}}, \bibinfo {author} {\bibfnamefont {S.}~\bibnamefont {Pang}}, \bibinfo
  {author} {\bibfnamefont {Y.}~\bibnamefont {Zhou}}, \ and\ \bibinfo {author}
  {\bibfnamefont {A.~N.}\ \bibnamefont {Jordan}},\ }\href {\doibase
  10.1103/PhysRevA.100.032104} {\bibfield  {journal} {\bibinfo  {journal}
  {Phys. Rev. A}\ }\textbf {\bibinfo {volume} {100}},\ \bibinfo {pages}
  {032104} (\bibinfo {year} {2019})}\BibitemShut {NoStop}%
\bibitem [{\citenamefont {Ge}\ \emph {et~al.}(2018)\citenamefont {Ge},
  \citenamefont {Jacobs}, \citenamefont {Eldredge}, \citenamefont {Gorshkov},\
  and\ \citenamefont {Foss-Feig}}]{ge2018distributed}%
  \BibitemOpen
  \bibfield  {author} {\bibinfo {author} {\bibfnamefont {W.}~\bibnamefont
  {Ge}}, \bibinfo {author} {\bibfnamefont {K.}~\bibnamefont {Jacobs}}, \bibinfo
  {author} {\bibfnamefont {Z.}~\bibnamefont {Eldredge}}, \bibinfo {author}
  {\bibfnamefont {A.~V.}\ \bibnamefont {Gorshkov}}, \ and\ \bibinfo {author}
  {\bibfnamefont {M.}~\bibnamefont {Foss-Feig}},\ }\href@noop {} {\bibfield
  {journal} {\bibinfo  {journal} {Physical Review Letters}\ }\textbf {\bibinfo
  {volume} {121}},\ \bibinfo {pages} {043604} (\bibinfo {year}
  {2018})}\BibitemShut {NoStop}%
\bibitem [{\citenamefont {Polino}\ \emph {et~al.}(2019)\citenamefont {Polino},
  \citenamefont {Riva}, \citenamefont {Valeri}, \citenamefont {Silvestri},
  \citenamefont {Corrielli}, \citenamefont {Crespi}, \citenamefont {Spagnolo},
  \citenamefont {Osellame},\ and\ \citenamefont {Sciarrino}}]{Polino19}%
  \BibitemOpen
  \bibfield  {author} {\bibinfo {author} {\bibfnamefont {E.}~\bibnamefont
  {Polino}}, \bibinfo {author} {\bibfnamefont {M.}~\bibnamefont {Riva}},
  \bibinfo {author} {\bibfnamefont {M.}~\bibnamefont {Valeri}}, \bibinfo
  {author} {\bibfnamefont {R.}~\bibnamefont {Silvestri}}, \bibinfo {author}
  {\bibfnamefont {G.}~\bibnamefont {Corrielli}}, \bibinfo {author}
  {\bibfnamefont {A.}~\bibnamefont {Crespi}}, \bibinfo {author} {\bibfnamefont
  {N.}~\bibnamefont {Spagnolo}}, \bibinfo {author} {\bibfnamefont
  {R.}~\bibnamefont {Osellame}}, \ and\ \bibinfo {author} {\bibfnamefont
  {F.}~\bibnamefont {Sciarrino}},\ }\href@noop {} {\bibfield  {journal}
  {\bibinfo  {journal} {Optica}\ }\textbf {\bibinfo {volume} {6}},\ \bibinfo
  {pages} {288} (\bibinfo {year} {2019})}\BibitemShut {NoStop}%
\bibitem [{\citenamefont {Macr\`{\i}}\ \emph {et~al.}(2016)\citenamefont
  {Macr\`{\i}}, \citenamefont {Smerzi},\ and\ \citenamefont
  {Pezz\`e}}]{macri2016loschmidt}%
  \BibitemOpen
  \bibfield  {author} {\bibinfo {author} {\bibfnamefont {T.}~\bibnamefont
  {Macr\`{\i}}}, \bibinfo {author} {\bibfnamefont {A.}~\bibnamefont {Smerzi}},
  \ and\ \bibinfo {author} {\bibfnamefont {L.}~\bibnamefont {Pezz\`e}},\ }\href
  {\doibase 10.1103/PhysRevA.94.010102} {\bibfield  {journal} {\bibinfo
  {journal} {Phys. Rev. A}\ }\textbf {\bibinfo {volume} {94}},\ \bibinfo
  {pages} {010102} (\bibinfo {year} {2016})}\BibitemShut {NoStop}%
\bibitem [{\citenamefont {Dirac}(1981)}]{dirac1981principles}%
  \BibitemOpen
  \bibfield  {author} {\bibinfo {author} {\bibfnamefont {P.~A.~M.}\
  \bibnamefont {Dirac}},\ }\href@noop {} {\emph {\bibinfo {title} {The
  principles of quantum mechanics}}},\ \bibinfo {number} {27}\ (\bibinfo
  {publisher} {Oxford university press},\ \bibinfo {year} {1981})\BibitemShut
  {NoStop}%
\bibitem [{\citenamefont {Helstrom}(1976)}]{helstromquantum}%
  \BibitemOpen
  \bibfield  {author} {\bibinfo {author} {\bibfnamefont {C.~W.}\ \bibnamefont
  {Helstrom}},\ }\href@noop {} {\emph {\bibinfo {title} {Quantum Detection and
  Estimation Theory}}}\ (\bibinfo  {publisher} {Academic Press},\ \bibinfo
  {year} {1976})\BibitemShut {NoStop}%
\bibitem [{\citenamefont {Paris}(2009)}]{paris2009quantum}%
  \BibitemOpen
  \bibfield  {author} {\bibinfo {author} {\bibfnamefont {M.~G.}\ \bibnamefont
  {Paris}},\ }\href@noop {} {\bibfield  {journal} {\bibinfo  {journal}
  {International Journal of Quantum Information}\ }\textbf {\bibinfo {volume}
  {7}},\ \bibinfo {pages} {125} (\bibinfo {year} {2009})}\BibitemShut {NoStop}%
\bibitem [{\citenamefont {Kay}(1993)}]{kay1993fundamentals}%
  \BibitemOpen
  \bibfield  {author} {\bibinfo {author} {\bibfnamefont {S.~M.}\ \bibnamefont
  {Kay}},\ }\href@noop {} {\emph {\bibinfo {title} {Fundamentals of Statistical
  Signal Processing}}}\ (\bibinfo  {publisher} {Upper Saddle River:Prentice
  Hall PTR},\ \bibinfo {year} {1993})\BibitemShut {NoStop}%
\bibitem [{\citenamefont {Luo}\ \emph {et~al.}(2017)\citenamefont {Luo},
  \citenamefont {Zou}, \citenamefont {Wu}, \citenamefont {Liu}, \citenamefont
  {Han}, \citenamefont {Tey},\ and\ \citenamefont
  {You}}]{luo2017deterministic}%
  \BibitemOpen
  \bibfield  {author} {\bibinfo {author} {\bibfnamefont {X.-Y.}\ \bibnamefont
  {Luo}}, \bibinfo {author} {\bibfnamefont {Y.-Q.}\ \bibnamefont {Zou}},
  \bibinfo {author} {\bibfnamefont {L.-N.}\ \bibnamefont {Wu}}, \bibinfo
  {author} {\bibfnamefont {Q.}~\bibnamefont {Liu}}, \bibinfo {author}
  {\bibfnamefont {M.-F.}\ \bibnamefont {Han}}, \bibinfo {author} {\bibfnamefont
  {M.~K.}\ \bibnamefont {Tey}}, \ and\ \bibinfo {author} {\bibfnamefont
  {L.}~\bibnamefont {You}},\ }\href@noop {} {\bibfield  {journal} {\bibinfo
  {journal} {Science}\ }\textbf {\bibinfo {volume} {355}},\ \bibinfo {pages}
  {620} (\bibinfo {year} {2017})}\BibitemShut {NoStop}%
\bibitem [{\citenamefont {Zou}\ \emph {et~al.}(2018)\citenamefont {Zou},
  \citenamefont {Wu}, \citenamefont {Liu}, \citenamefont {Luo}, \citenamefont
  {Guo}, \citenamefont {Cao}, \citenamefont {Tey},\ and\ \citenamefont
  {You}}]{zou2018beating}%
  \BibitemOpen
  \bibfield  {author} {\bibinfo {author} {\bibfnamefont {Y.-Q.}\ \bibnamefont
  {Zou}}, \bibinfo {author} {\bibfnamefont {L.-N.}\ \bibnamefont {Wu}},
  \bibinfo {author} {\bibfnamefont {Q.}~\bibnamefont {Liu}}, \bibinfo {author}
  {\bibfnamefont {X.-Y.}\ \bibnamefont {Luo}}, \bibinfo {author} {\bibfnamefont
  {S.-F.}\ \bibnamefont {Guo}}, \bibinfo {author} {\bibfnamefont {J.-H.}\
  \bibnamefont {Cao}}, \bibinfo {author} {\bibfnamefont {M.~K.}\ \bibnamefont
  {Tey}}, \ and\ \bibinfo {author} {\bibfnamefont {L.}~\bibnamefont {You}},\
  }\href@noop {} {\bibfield  {journal} {\bibinfo  {journal} {Proceedings of the
  National Academy of Sciences}\ }\textbf {\bibinfo {volume} {115}},\ \bibinfo
  {pages} {638} (\bibinfo {year} {2018})}\BibitemShut {NoStop}%
\bibitem [{\citenamefont {Jarzyna}\ and\ \citenamefont
  {Demkowicz-Dobrza\ifmmode~\acute{n}\else
  \'{n}\fi{}ski}(2012)}]{jarzyna2012quantum}%
  \BibitemOpen
  \bibfield  {author} {\bibinfo {author} {\bibfnamefont {M.}~\bibnamefont
  {Jarzyna}}\ and\ \bibinfo {author} {\bibfnamefont {R.}~\bibnamefont
  {Demkowicz-Dobrza\ifmmode~\acute{n}\else \'{n}\fi{}ski}},\ }\href {\doibase
  10.1103/PhysRevA.85.011801} {\bibfield  {journal} {\bibinfo  {journal} {Phys.
  Rev. A}\ }\textbf {\bibinfo {volume} {85}},\ \bibinfo {pages} {011801}
  (\bibinfo {year} {2012})}\BibitemShut {NoStop}%
\bibitem [{\citenamefont {Hyllus}\ \emph {et~al.}(2010)\citenamefont {Hyllus},
  \citenamefont {Pezz{\'e}},\ and\ \citenamefont
  {Smerzi}}]{hyllus2010entanglement}%
  \BibitemOpen
  \bibfield  {author} {\bibinfo {author} {\bibfnamefont {P.}~\bibnamefont
  {Hyllus}}, \bibinfo {author} {\bibfnamefont {L.}~\bibnamefont {Pezz{\'e}}}, \
  and\ \bibinfo {author} {\bibfnamefont {A.}~\bibnamefont {Smerzi}},\
  }\href@noop {} {\bibfield  {journal} {\bibinfo  {journal} {Physical review
  letters}\ }\textbf {\bibinfo {volume} {105}},\ \bibinfo {pages} {120501}
  (\bibinfo {year} {2010})}\BibitemShut {NoStop}%
\bibitem [{\citenamefont {Pezz\`e}\ \emph {et~al.}(2015)\citenamefont
  {Pezz\`e}, \citenamefont {Hyllus},\ and\ \citenamefont
  {Smerzi}}]{pezze2015phase}%
  \BibitemOpen
  \bibfield  {author} {\bibinfo {author} {\bibfnamefont {L.}~\bibnamefont
  {Pezz\`e}}, \bibinfo {author} {\bibfnamefont {P.}~\bibnamefont {Hyllus}}, \
  and\ \bibinfo {author} {\bibfnamefont {A.}~\bibnamefont {Smerzi}},\ }\href
  {\doibase 10.1103/PhysRevA.91.032103} {\bibfield  {journal} {\bibinfo
  {journal} {Phys. Rev. A}\ }\textbf {\bibinfo {volume} {91}},\ \bibinfo
  {pages} {032103} (\bibinfo {year} {2015})}\BibitemShut {NoStop}%
\bibitem [{Var()}]{Varenna}%
  \BibitemOpen
  \href@noop {} {}\bibinfo {note} {L. Pezz\`{e} and A. Smerzi, \textit{Quantum
  theory of phase estimation} in G. M. Tino and M. A. Kasevich (Eds.),
  \textit{Atom Interferometry. Proceedings of the International School of
  Physics Enrico Fermi, Course 188, Varenna}, 691--741, IOS Press
  (2014).}\BibitemShut {Stop}%
\end{thebibliography}
%

\end{document}